\documentclass{pasj00}
\draft

\renewcommand{\vector}{\mathbf}

\newcommand{\varkappa}{\kappa}
\newcommand{\keV}{\ensuremath{\rm keV}}
\newcommand{\ergl}{\ensuremath{\rm ~erg \,s^{-1}}}

\newcommand{\arctg}{\arctan}
\newcommand{\tg}{\tan}
\newcommand{\cmc}{\ensuremath{{\rm cm^{-3}}}\,}
\newcommand{\mdot}{\ensuremath{\dot{m}}}
\newcommand{\xmm}{{\it XMM-Newton}}
\newcommand{\galex}{{\it GALEX}}
\newcommand{\hst}{{\it HST}}
\newcommand{\Msun}{\ensuremath{\rm M_\odot}}
\newcommand{\Ry}{\ensuremath{\rm Ry}}
\newcommand{\yr}{\ensuremath{\rm yr}}
\newcommand{\Myr}{\ensuremath{\rm Myr}}
\newcommand{\magdot}[1]{\ensuremath{^{\rm m}\!\!#1\,}}
\renewcommand{\div}{\ensuremath{-}}
\newcommand{\rad}{\ensuremath{\rm rad}}
\newcommand{\kms}{\ensuremath{\rm km \,s^{-1}}}

\begin{document}
\SetRunningHead{Abolmasov et al.}{Supercritical Funnel}
\title{Optically Thick Outflows of Supercritical Accretion Discs: Radiative
Diffusion Approach}
\author{P. \textsc{Abolmasov}$^1$, S. \textsc{Karpov}$^1$ \and Taro \textsc{Kotani}$^{2}$}
\affil{$^1$Special Astrophysical Observatory, Nizhnij Arkhyz 369167, Russia}
\affil{$^2$Dept. of Physics, Tokyo Tech, 2-12-1 O-okayama, Tokyo 152-8551}
\email{pasha@sao.ru}       

\maketitle

\begin{abstract}
  Highly supercritical accretion discs are probable sources of dense
  optically thick axisymmetric winds. 
We introduce a new approach
  based on diffusion approximation radiative transfer in a funnel
  geometry and obtain an analytical solution for the
  energy density distribution inside the wind 
  assuming that all the mass, momentum
  and energy are injected well inside the spherization radius. 
This allows to
  derive the spectrum of emergent emission for various inclination angles.
We show that self-irradiation effects play an important role
  altering the temperature of the outcoming radiation by about 20\%
  and the apparent X-ray luminosity by a factor of $2\div 3$.
The model has been successfully applied to two ULXs.
The basic properties of the high ionization HII-regions found around
some ULXs are also easily reproduced in our assumptions. 
\end{abstract}

\section{Introduction}\label{sec:intro}

Processes of gas accretion onto compact objects are studied since 60s
when first X-ray sources were discovered
\citep{accretion_power}. Among several works describing the details of
disc accretion in binary systems \citet{ss73} was the most
successfull. Their ``standard disc'' is still widely used to describe
the thermal component of X-ray spectra of X-ray binaries. Standard
disc model was worked out in several considerations such as high
optical thickness, low geometrical thickness of the disc, etc. Among
these assumptions one was that the power released in the accretion
process does not exceed the Eddington luminosity.
 This case is usually
called subcritical accretion.

However, considering various phenomena such as growth of
supermassive black holes and nova outbursts leads to the problem of supercritical
accretion. It has been extensively investigated since 1980s when Abramowicz
proposed it as a power source for active galactic nuclei
\citep{agn_tori}. Presently super-Eddington accretion is often applied to
explain the observational properties of Ultraluminous X-ray Sources (ULXs,  see
\citet{roberts_review} for observational review) that
are believed to be high-mass X-ray binaries with black holes accreting on
thermal time scale.


The first, outflow-dominated model of supercritical accretion, has been
developed by Shakura and Sunyaev in the paper mentioned above.
Authors assumed that all the inflowing gas above the critical
accretion rate is being ejected from the disc in the form of a wind. Another version
of supercritical regime is based on the relaxation of the locality condition in
``standard model'' which leads to  advective slim discs \citep{agn_tori} or
Polish doughnuts \citep{abram78,kozl78,abram2004}. In reality, both processes
work simultaneously.

A comprehensive model taking into account both for
advection and ejection was recently developed by \citet{poutanen}.
Authors consider the structure of the disc in radial direction and
estimate three characteristic temperatures relevant for the outcoming
radiation (inner disc temperature, temperature at spherization radius
and effective temperature of the wind photosphere) but do not study
the processes of radiation transfer in the wind and do not calculate
the outcoming spectra.

Obviously, at high accretion rates the observational properties of the
accretion flow are governed mostly by radiative transfer in the
outflowing wind. 
The observational appearance of the pseudo-photosphere of the
optically-thick wind of a supercritical accretion disc was considered
by \citet{nishi07}. The authors however assumed the optically-thick
part of the wind fully adiabatic not taking into account radiative
energy transfer that is expected to be important in the flow (see
section \ref{sec:diff}). 

Currently, we have at least one example of a persistent supercritical accretor
in our Galaxy -- the SS433 system, where most of accreting gas is being
lost in a wind.
Numerical simulations of this system
partially support the outflow-dominated scenario. 
However, \citet{okuda} states that his thorough 2D simulation fails to
reproduce the outflow rate and jet collimation in SS433.
In \citet{ohsuga2005} a supercritical accretor
accreting at $ 10^3$ Eddington rate
appears a bright (about $10^{38}\ergl$) hard X-ray source if viewed
edge-on. \citet{ohsuga2005} and \citet{heinz} calculate the structure
and emergent spectra considering only the inner parts of the
flow ($R \le 500 R_G$) not taking into account that the region
considered is coated by accreting and outflowing matter both being
optically thick. The outcoming spectra will strikingly differ from
those calculated by \citet{heinz}.


Optical and radio \citep{blundell2001} 
observations allow to measure the mass ejection rate
in the relatively slow ($1000\div 2000\,\kms$) accretion disc wind seen in optical emission
and absorption lines \citep{ss2004} as well as the mass loss rate in
the collimated mildly-relativistic jets launched along the disc
axis. Infrared excess was used by \citet{sklov81} and \citet{vdh81} to
estimate the mass ejection rate from SS433. No direct mass transfer
rate measurements were ever made though it is usually supposed that it
could not be much higher than the mass ejection rate $\dot{M}_{ej}
\sim 10^{-4}\, \Msun\, \yr^{-1}$ \citep{ss2004}. 

The details of the processes  in the supercritical disc itself are 
unclear mainly due to the strong absorption in the wind. In fact even the
photosphere of the wind is difficult to study due to high
interstellar absorption (about $8\magdot{\,}$) making it
impossible with the contemporary instrumentation to trace the Spectral
Energy Distribution (SED) in the far ultraviolet (UV) spectral range were most of the
radiation is expected to be emitted.
 The only and yet crude estimate of the blackbody temperature 
of SS433 was made by \citet{dolan} and probably corresponds to the
photosphere of the wind.
Their measurements are consistent with a $(2\div7) \times 10^4\,\rm K$
blackbody source.

Both observations of SS433 \citep{ss2004} and numerical simulations
\citep{okuda,ohsuga2005} support the conception that is traditionally called
``supercritical funnel''. It assumes that nearly all the matter accreted is being
ejected in a form of a slow, roughly virial at spherization radius, dense
wind. Due to centrifugal barrier two conical avoidance sectors with
half-opening angles of $\theta_f \sim 30^{\circ}$ \citep{EGK} are formed along
the accretion disc symmetry axis, filled with rarefied hot gas, which may be
accelerated and collimated to form relativistic jets. In the
inner parts of the wind this gas also can form a pseudo-photosphere
(``funnel bottom'')
at $R_{in} \sim 10^9 \rm cm$  \citep{ss2004}. This radius is calculated in
the consideration all the material ejected in the relativistic jets is
uniformly distributed with respect to the polar angle. In case of any
inhomogeneity of the jet material the inner radius becomes lower. 
The visible part of the wind may be therefore divided into three
parts: funnel wall photosphere (or ``photocone''), the outer
photosphere and the inner photosphere inside the funnel. 

Another class of objects supposed to be supercritical accretors are
extragalactic Ultraluminous X-ray sources (ULXs).
\citet{katz86} supposed that an object similar to SS433 seen at low inclination angles
can appear a bright X-ray source with super-Eddington apparent luminosity. 
ULXs were discovered about that time by {\it Einstein}
(see \citet{fabbiano88} and references therein). Though the nature of
these sources remains unclear
they are good candidates for the objects predicted by
Katz \citep{poutanen}.  

The question about how representative is SS433 among the binary
systems in the supercritical accretion regime in
the observed Universe is difficult to answer.
If one considers mass transfer in thermal timescale in a black
hole high mass X-ray binary,
the most relevant is the mass of the secondary that determines the
timescale and hence the scaling for the mass accretion rate.
It may be shown that in assumption the radius of the secondary scales
with its mass as $R \propto M^{1/2}$ and the black hole mass is close to
$10~M_\odot$ dimensionless thermal-timescale mass transfer rate
depends on the secondary mass as:

$$
\mdot \simeq \left(\frac{M_2}{M_\odot}\right)^{5/2}.
$$

For the case of SS433 ($M_2 \simeq 20\,M_\odot$) our estimate predicts
$\mdot \sim 2000$, similar to the observed value
\footnote{Acutally,
  there are no direct measurements of the mass transfer rate for
  SS433. However, the observed stability of the orbital period and
  evolution-time considerations exclude significantly higher mass
  transfer rates \citep{ss2004,gorss}.}.
 If the donor star is highly
evolved, accretion rate is higher and less stable and the phase
itself is shorter.
Unfortunately, initial binary mass ratio distribution for massive
  stars is poorly known but there are indications that mass ratios
  close to 1 are much more probable \citep{lucy,KoFr2007}.
The thermal timescale mass transfer rates in high-mass X-ray binaries 
are therefore likely to be highly
  supercritical, $\mdot \sim 100\div 10^4$. Lower yet supercritical rates
  may take place in case of wind accretion for example in WR+BH binaries
  \citep{ic10_bauer}.
Those are likely to form a certain sub-sample of evolved ULXs with
  moderately supercritical accretion rates.


The paper is organized as follows. In the next section we 
construct a simple analytical model describing
the internal structure of an optically thick wind flow that is likely
to develop in the case of very high mass ejection (and accretion) rate
\mdot.
Section \ref{sec:irrad} is devoted to effects of self-irradiation that are likely to play very
important role in our funnel solution. 
In section  \ref{sec:sp} we consider the emergent SEDs for arbitrary
inclination taking into account self-occultation effects.
In section \ref{sec:obs} we 
test it for two sets of publicly available X-ray data. 
We discuss the implications of the model and its early testing in
section \ref{sec:disc}.

\section{Structure of Supercritical Disc Winds}\label{sec:anal}

In this section we study the spatial structure of the radiation density field
inside the wind. We will use the following set of assumptions (their reliability
will be discussed below in Section~\ref{sec:disc}):
{\it (i)} the flow is axisymmetric (also symmetric with respect to
  the disc plane) and stationary,
{\it (ii)} the flow is optically thick to true absorption,
{\it (iii)} the velocity and density fields are not affected by energy
  and entropy transfer (i. e. we consider the wind already accelerated
  or accelerating/decelerating  with a given power-law dependence
  on radius),
{\it (iv)} all the motions are purely radial and non-relativistic,
{\it (v)} the inner surface of the funnel, the funnel bottom and the
  wind photosphere are considered locally blackbody sources and
{\it (vi)} for certainty we suggest the temperature  of the bottom equal
  to the starting temperature of the walls.
We simplified the picture somehow suggesting the mass is loaded in the
center of symmetry.
This assumption is violated in the inner parts of the wind.

Effects of special relativity (namely, Doppler boosts) may be
approximately accounted for during the calculation of outcoming
spectra. We also discuss the possible influence of relativistic
effects in sections \ref{sec:sp} and \ref{sec:trep}. 

We assume that fraction $K \lesssim 1$ of the accreting mass is ejected
in the wind. Because mass ejection rate is more relevant (see below) we will use the
 dimensionless notation
$\mdot = \dot{M} / \dot{M}_{cr}$ for the {\bf mass ejection rate} where 
\begin{equation}
\dot{M}_{cr} = \frac{48\pi GM }{c\kappa_T} \simeq 3 \times 10^{-7} 
M_{10} \ \Msun \, \yr^{-1}
\end{equation}
is the critical Eddington accretion rate as intruduced by
\citet{poutanen}. $M_{10}$ here is the accretor mass in $10$\Msun\ units.
 Mass accretion is characterised by $\mdot / K \sim
\mdot$, we assume $K=1$ everywhere below bearing in mind that $K<1$
results in a hotter and more luminous source but requires higher
accretion rate. The outcoming spectra depend mostly on the mass
ejection rate. Mass accretion rate affects only the luminosity
injected at the inner boundary in a logarythmic way. To achive a 50\%
change in luminosity (and about 12\% change in temperature) for $\mdot
= 100$ one should assume 90\% of the accreting material accreted by
the black hole ($K = 0.1$). 

There are physical differences between the regime of highly
supercritical accretion that we consider here and
moderately supercritical accretion ($\mdot \sim 1\div 10$). 
At $\mdot \sim 10$ the flow becomes translucent and
relativistic and some of our approximations are
violated.

For radial coordinate normalization we will use the ``spherization
radius'' defined as:
\begin{equation}
R_{sph} = \frac{3\dot{M}\varkappa}{\Omega_f c} = \frac{18}{\cos\theta_f} R_G \mdot \mbox{.}
\end{equation}
\noindent
$R_G = 2GM / c^2$ here is the gravitational (Schwarzschild) radius of the
accretor. 
This value is proportional to the spherization radii used by \citet{ss73}
\begin{equation}
R_{sph}^{(SS)} = \frac{3\dot{M}\varkappa}{8\pi c}
\end{equation}
\noindent
and by \citet{poutanen}
\begin{equation}
R_{sph}^{(P)} = 5/3 \mdot \mbox{,}
\end{equation}
\noindent
but has also non-trivial dependence on geometry. $\Omega_f = 4\pi
\cos\theta_f$ is the solid angle of the wind. 
Normalized radial coordinates
will be hereafter denoted by small letters $r$ in contrast to capital
$R$ reserved for physical distances.

We assume fixed geometry for the wind funnel, i.e. fixed half-opening angle
$\theta_f$ independent of the accretion/ejection rate. Note the difference with funnels in
Polish doughnuts \citep{abram2004} where
\begin{equation}
\theta_f \propto \sqrt{r_{in}/r_{out}} \propto e^{-0.5L/L_{Edd}} \propto
\mdot^{-0.3},
\end{equation}
\noindent
defined by equiponetial surfaces. In contrast, when the wind is
launched from a thin but supercritical disk at a certain radius and its
velocity in the frame comoving with the disk is at any given starting
radius proportional to the virial velocity ($v = \xi v_K$ where $v_K$ is
the Keplerian velocity) and normal to the disk surface, the half-opening angle
will be a function of $\xi$ only \citep{poutanen}:
\begin{equation}
\theta_f = \arctg \left(1/\sqrt{\xi^2-1} \right)  \mbox{.}
\end{equation}
Radial velocity of the wind may be uniform ($v=const$) in the case of large
initial speed or virial ($v \propto R^{-1/2}$) in the parabolic case when the initial
velocity is close to the escape velocity from the wind acceleration
region. We adopt a generalized self-similar scaling

\begin{equation}\label{E:velty}
v=\frac{1}{6}\sqrt{\cos\theta_f} c \mdot^{-1/2} r^{\alpha}
\end{equation}
\noindent
with approximately virial value at the spherization radius. 
$\alpha=0$ corresponds to a constant velocity wind with no
acceleration. In the parabolic case, when the velocity is proportional
to the virial at any given radius, $\alpha= -0.5$. 

Outside the spherization radius the gas density in the wind scales as
\begin{equation}
n \propto r^{-(2+\alpha)}
\end{equation}
\noindent
and vanishes (or, at least, drops by several orders of magnitude) inside the
funnel. Inside the spherisation
radius deviations from this law are expected and mass ejection rate
should depend on radius roughly as $\dot{m}_{eff} \propto r$, and hence
\begin{equation}
n \propto r^{(-1+\alpha)} \mbox{.}
\end{equation}
\noindent
We briefly analyse the consequences of this difference in density slope in section~\ref{sec:disc}.

\subsection{Diffusion Equation}\label{sec:diff}

The main equation governing energy transfer may be derived from
the energy conservation and Fick's laws for the
thermal energy flux
\begin{equation}
\vector{q} = - D \nabla u \mbox{,}
\end{equation}
where $D = c / (3\varkappa \rho)$ is the diffusion coefficient and $u$ is
energy density. The first law of thermodynamics
\begin{equation}
  d\left(\frac{u}{n}\right) = T ds + \frac{p}{n^2}dn
\end{equation}
\begin{equation}
  du = nT ds + \frac{p+u}{n}dn
\end{equation}
where
\begin{equation}
  T ds = (\nabla\vector{q}) dt
\end{equation}
\begin{equation}
  du/dt = \partial_t u + (\vector{v} \nabla) u
\end{equation}
\begin{equation}
  dn/dt = - n (\nabla\vector{v})
\end{equation}
may be rewritten in the generic form
\begin{equation}\label{E:gener}
  \partial_t u + (\vector{v} \nabla) u - \nabla\left( D \nabla u\right)
  - \gamma u (\nabla\vector{v})=0
\end{equation}
\noindent

If the specific heat ratio $\gamma$ does not depend heavily on other
parameters, this may be rewritten for an axisymmetric system in terms of
enthalpy density $h = \gamma u$ as
\begin{equation}\label{E:ueq}
\begin{array}{l}
  \frac{1}{v} \partial_t u + \partial_r u - \frac{1}{r^{2+\alpha}}
  \partial_r \left( r^{4+\alpha} \partial_r u\right) - \\
 \qquad{} - \left( L_\theta - \frac{\gamma (2+\alpha) }{r}\right) u =0\\
\end{array}
\end{equation}
where
\begin{equation}
  L_\theta = \frac{1}{\sin\theta} \partial_\theta \left(
    \sin\theta \partial_\theta ~~ \right)
\end{equation}
is the polar part of the Laplace operator, and the second term in brackets
accounts for adiabatic losses. Equation (\ref{E:ueq}) is linear hence
one may separate the variables. Eigenfunctions will depend on
$\cos\theta$ as linear combinations of Legendre polynomials $P_k$ and
Legendre functions of the second kind $Q_k$ \citep{Q_weisstein}.

In the stationary case
\begin{equation}
u_k=\left( P_k(\cos \theta) + a_k Q_k(\cos \theta)  \right) R_k(r)
\end{equation}
where $R_k$ is the solution of the equation
\begin{equation}\label{E:req}
\begin{array}{l}
  - \frac{1}{r^{2+\alpha}} e^{-1/r} \partial_r \left( r^{4+\alpha}
    e^{1/r} \partial_r u\right) - \\
\qquad{} - \left( k (k+1) - \frac{\gamma (2+\alpha)
    }{r}\right) u =0 \\
\end{array}
\end{equation}

\begin{figure}
\FigureFile(\columnwidth,\texthight){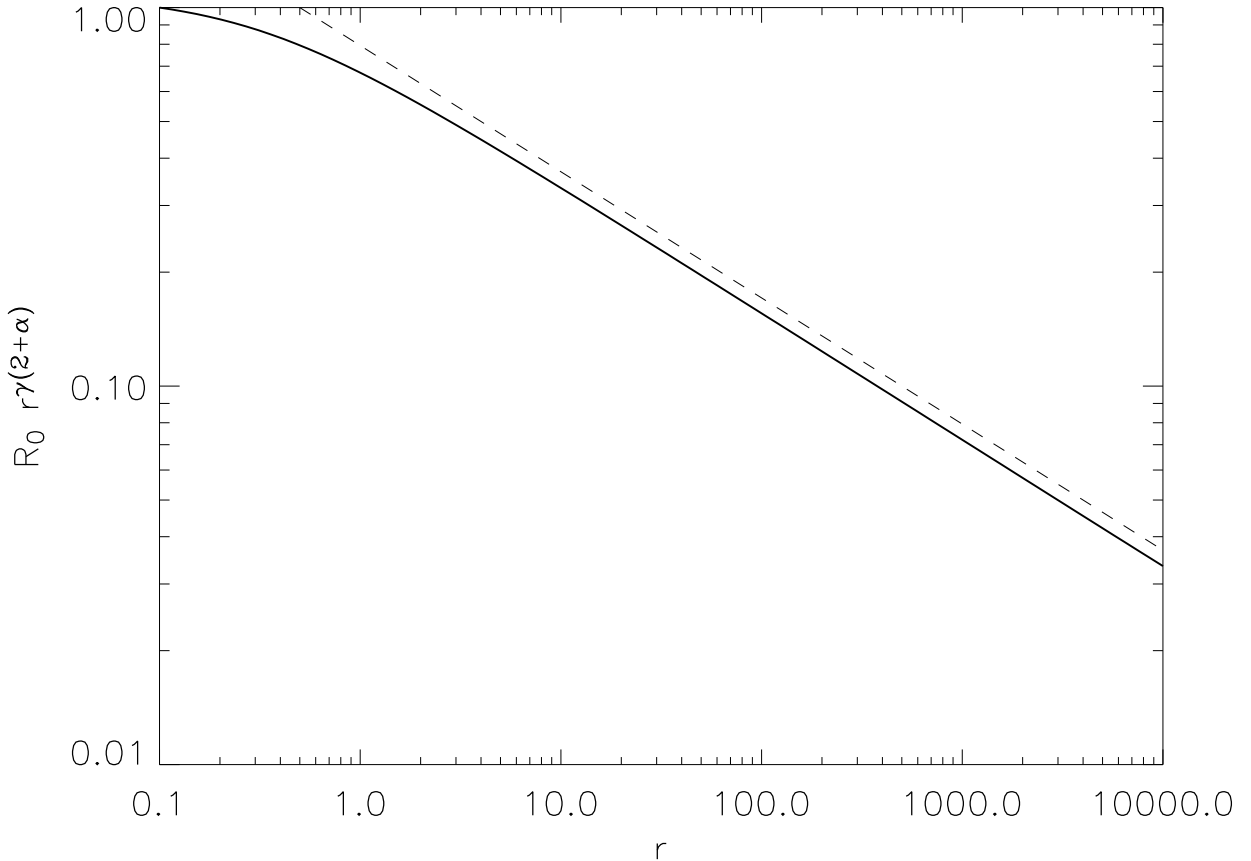}
  \caption{Radial dependence for the zeroth-order solution for $\alpha=0$ and
    $\gamma=4/3$ (solid line) normalized by the low-$r$ asymptotics. 
High-$r$ power-law asymptotics $u\propto r^{-(2+\alpha)}$ is shown by
    a dashed curve.
\label{fig:eir}}
\end{figure}

For the zeroth-order ($k=0$) solution asymptotics may be derived both for
large distances $r \gg 1$ where diffusion dominates, $e^{-1/r} \simeq 1$ and
$u \propto r^{-(3+\alpha)}$, and for very low radii where adiabatic losses
prevail and $u \propto r^{-\gamma (2+\alpha)}$. Figure ~\ref{fig:eir} shows the
exact numerical solution for the radial eigenfunction $R_0$ for $\alpha = 0$
and $\gamma=4/3$. 
It may be rather well (with accuracy better than 2\%) approximated by the
following function
\begin{equation}\label{E:Y}
  Y(r) = \left( 1 - e^{-\frac{1}{2+\alpha} \frac{1}{r}}\right)^{1 - (\gamma-1) (2+\alpha)} r^{-\gamma (2+\alpha)}
\end{equation}

\subsection{Boundary Conditions}

For a system with equatorial plane symmetry the solution
will contain only even eigenfunctions. If we choose the inner radius $R_{in}$
well below the spherization radius, equation~(\ref{E:req}) becomes a
first-order one, and is governed by a single parameter which we assume to be
the initial energy density. The latter may be estimated from the total central
source luminosity $L$ for the advective flow as
\begin{equation}\label{E:uin}
u_{in} \sim \frac{L}{\Omega R^2_{in} v(r_{in}) Y(r_{in})} 
\end{equation}

Luminosity of a supercritical disk in different models depends
logarithmically on the dimensionless mass accretion rate \citep{poutanen,ss73}
as
\begin{equation}\label{E:lum}
 \begin{array}{l}
   L = L_{Edd} \left( 1 + 0.6 \ln (\mdot /K) \right) \simeq \\
    \qquad{} \simeq 1.5 \times 10^{39} \frac{1.7}{1+X} M_{10} \left( 2.8 + 0.6 \ln
    \mdot_3 \right) \ergl ,\\
 \end{array}
\end{equation}
\noindent
where $\mdot_3$ is the mass ejection rate in $10^3 \dot{M}_{cr}$, $X$
is hydrogen mass fraction in the accreting gas, $M_{10}$ is the
accretor mass in $10\,\Msun$ units.
Temperature of the flow at the spherization radius is
$T \sim 0.1 \mdot_3^{-3/8} ~\keV$ while the gas density is
$n \sim 4 \times 10^{16} \mdot_3^{-1/2}\rm cm^{-3}$, which leads to a rather high radiation to gas
pressure ratio:
\begin{equation}
\beta \simeq \frac{aT^3}{n} \simeq 500 \mdot_3^{-5/8} 
\end{equation}
\noindent
For accretion rates that appear during thermal-timescale mass transfer
$ 10 < \mdot < 10^5$ (see also sections ~\ref{sec:intro} and ~\ref{sec:disc})
radiation pressure dominates in the advective inner parts of the wind.
That allows to equate $\gamma=4/3$ anywhere in the flow because the
wind is radiation-pressure dominated in the regions where $\gamma$ is relevant.

\begin{figure}
\center{
\FigureFile(\columnwidth,\texthight){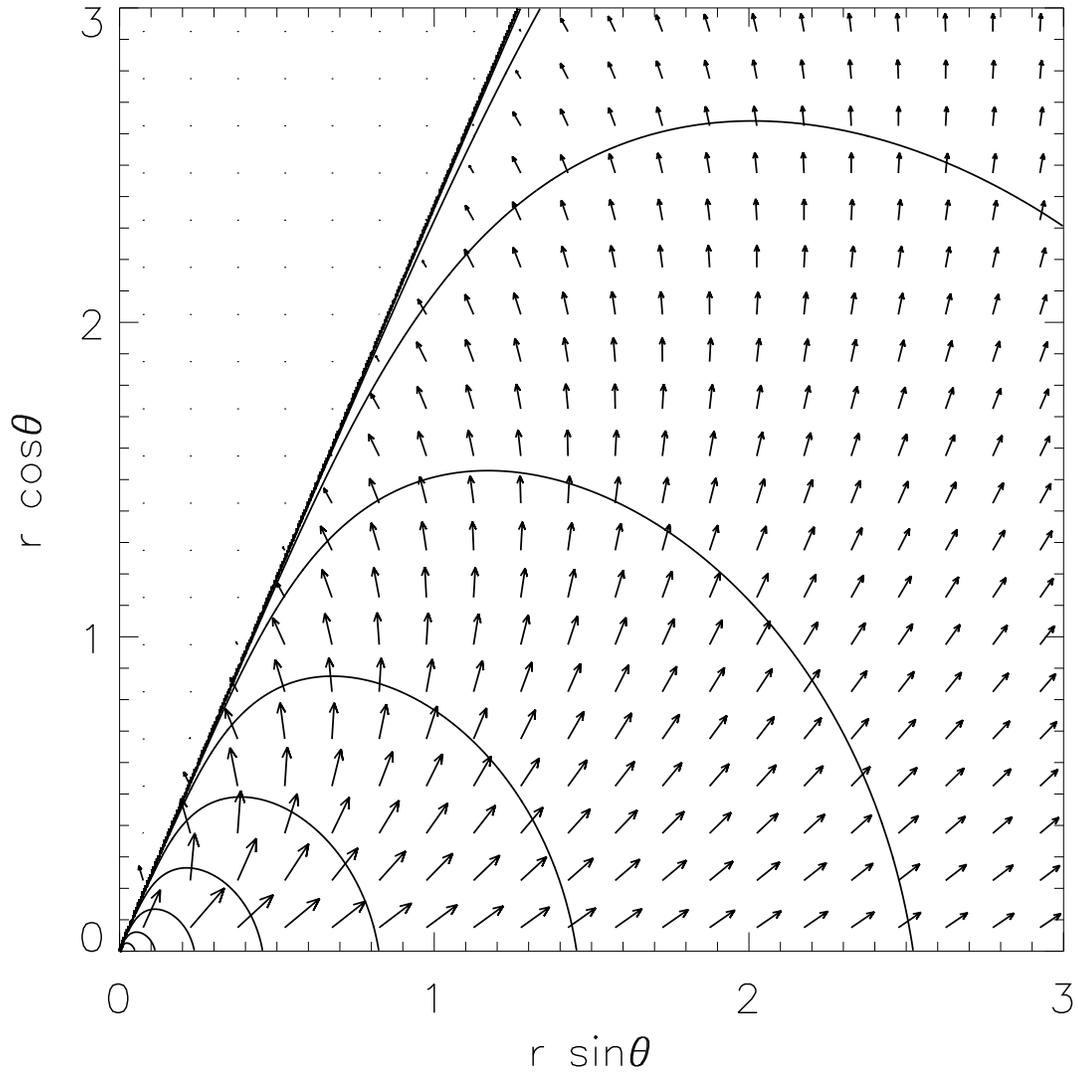}
\caption{Two-dimensional distribution of energy density in the
  optically thick wind. 
Thin solid lines correspond to logarythmically spaced (by one order of
  magnitude) constant energy density levels. 
 $\theta_f = 0.4\,\rad \simeq 23^\circ$, $\alpha=0$.
  Inner funnel wall is shown by a thick solid line. Arrows are
  oriented in the energy flux direction, their length proportional to
  the energy flux logarythm.}\label{fig:2d}
}
\end{figure}

Energy flux is directed radially in the inner advective parts of the
wind and is practically normal to the walls near the funnel wall
surface in the outer parts. In figure~\ref{fig:2d} we show the
two-dimesional distribution of energy density and energy flux vector
field for our solution.
The exact angular dependence of the  energy influx at the inner boundary is not
very important as it is mixed near the spherization radius, so we will
restrict ourselves to the zeroth angular harmonic solution.


Funnel wall cools efficiently, therefore a simple boundary condition
\begin{equation}
  u(r,\theta_f) = 0
\end{equation}
\noindent
may be adopted. The following zeroth-order solution satisfies it and decays
fast enough at infinity to account for the photon escape through the funnel
walls
\begin{equation}\label{E:sol0}
  u = u_0 \times \ln \left( \frac{1-\cos\theta}{1-\cos\theta_f} \times
    \frac{1+\cos\theta_f}{1+\cos\theta} \right) Y(r)
\end{equation}
\noindent
$u_0$ normalization may be derived from the total
luminosity (more accurately than in (\ref{E:uin})) of the disk as

\begin{equation}
u_0 \simeq \frac{L}{\Omega^\prime R_{in}^2 v(r_{in}) Y(r_{in})}
\end{equation}
\noindent
Here we define $\Omega^\prime =- 8 \pi \ln \sin
\theta_f $. $Y(r_{in}) \simeq r_{in}^{-\gamma (2+\alpha)}$ for $r_{in}
\ll 1$. 

\subsection{Surface Effective Temperature Distribution}

Below we consider funnel walls, bottom and outer photospheres as blackbody
sources with the temperature determined by the 
energy flux normal component:
\begin{equation}
  F_n = \sigma_B T^4.
\end{equation}
For the funnel walls it is equal to the latitudal component:
\begin{equation}
  F_\theta = \frac{2 L}{\Omega^\prime \sin^2\theta_f R_{sph}^2}
 \frac{1}{r_{in}^{2+\alpha} Y(r_{in})} r^{1+\alpha} Y(r),
\end{equation}
\noindent
and in the case of outer photosphere
\begin{equation}
\begin{array}{l}
F_r =  - D \partial_r u = \frac{(3+\alpha) L}{\Omega^\prime R_{sph}^2}
\frac{1}{r_{in}^{2+\alpha} Y(r_{in})} r^{1+\alpha} Y(r)\times \\
\qquad{} \times \ln \left( \frac{1-\cos\theta}{1-\cos\theta_f} \times
    \frac{1+\cos\theta_f}{1+\cos\theta} \right) .\\
\end{array}
\end{equation}

Here $r=r_{out}$ should be substituted for the outer photosphere. 

Note that the fluxes are of the same order and their ratio does not
depend on the radial coordinate.

For the funnel walls one obtains:
\begin{equation}\label{E:wallT}
\begin{array}{l}
T_{wall} = 0.038 \left( \Omega^\prime \tg^2\theta_f
r_{in}^{2+\alpha}
 Y(r_{in}) \right)^{-1/4} \times\\
\times \left(2.8 + 0.6\ln \mdot_3\right)^{1/4}\mdot_3^{-1/2}
M_{10}^{-1/4}  \left( r^{1+\alpha} Y(r)\right)^{1/4} ~\keV\\
\end{array}
\end{equation}
\noindent
where $M_{10}$ is the mass of the accretor in $10\,\Msun$ units. Effective
temperature scales with radius  (for $\alpha =0$) as
\begin{equation}
  T_{wall} \propto r^{-5/12}
\end{equation}
\noindent
at lower radii and as
\begin{equation}
  T_{wall} \propto r^{-1/2}
\end{equation}
\noindent
in the outer parts of the funnel. 
Temperature slope depends very weakly on all the parameters and is
similar to the value $p=1/2$ (if $T \propto r^{-p}$) characteristic for slim discs
\citep{abram2004} and supercritical accretion discs with massive
outflows. 
{\it p-free} multi-blackbody spectra are characterised by a power-law
intermediate asymptotics with photon index $\Gamma=2/p-2$. Many
ULXs have power-law spectra with $\Gamma \simeq 2$ close to the
expected in all these models.
The similarity makes it very difficult to distinguish
between the various supercritical accretion disc models by their X-ray
spectra only.

We assume the temperature of the funnel bottom equal to the effective
temperature of the adjecent wall surface
\begin{equation}\label{E:botT}
\begin{array}{l}
T_{bot} = 0.038 \left( \Omega^\prime \tg^2\theta_f
r_{in} \right)^{-1/4} \\ 
\qquad{} \left(2.8 + 0.6\ln \mdot_3\right)^{1/4} \mdot_3^{-1/2}
M_{10}^{-1/4}  \,\keV\\
\end{array}
\end{equation}
\noindent
Minimal inner radius $R_{in}$ may be estimated as the last stable orbit radius
devided by $\cos\theta_f$ as $r_{in} = 1/6\mdot$ in $R_{sph}$ units. 
Temperatures characteristic for the high-energy cut-offs in ULX
spectra are about one \keV\ and require $r_{in} \gtrsim \mdot^{-1}$. 
It is clear that in the framework of our model the inner parts of the funnel should be
practically transparent.
 If we make another assumption that
the soft excess in ULX spectra corresponds to $T \sim 0.1\div
0.2~\keV$ temperature at the spherisation radius than accretion rates
required are $\mdot \sim 100$. In section ~\ref{sec:obs} we fit real
X-ray spectra with our model obtaining similar results ($r_{in} \sim
10^{-3}$ and $\mdot \sim 100$) for two ULXs. 


Effective temperature of the photosphere may be found in a similar
way. The outer photosphere radius may be estimated as
\begin{equation}
  r_{out} \simeq  \left( \frac{ 2 \Omega_f \sqrt{\mdot} }{
  (1+\alpha)\sqrt{\cos\theta_f}} \right)^{1/(1+\alpha)}
\end{equation}
\noindent
in $R_{sph}$ units. The photosphere is significantly non-plane-parallel, but
its luminosity and spectral energy distribution may be roughly
estimated
as:

\begin{equation}\label{E:phoL}
  L_{ph} \simeq \frac{(3+\alpha)
  (2+\alpha)^{(2+\alpha)(\gamma-1)-1}}{r_{in}^{2+\alpha} Y(r_{in})}
  \cos^2\theta_f L
\end{equation}
\begin{equation}\label{E:photemp}
\begin{array}{l}
  T_{ph} \simeq 0.0048 f_\alpha 
 \left( r_{in}^{2+\alpha} Y(r_{in})\right)^{-1/4}
 \sqrt{\cos\theta_f} \times \\
\qquad{} \times M_{10}^{-1/4} \left(2.8 + 0.6\ln
 \mdot_3\right)^{1/4} \times \\
\qquad{} \times \mdot_3^{-3/ 4(1+\alpha) } 
 \left(\ln \left( \frac{1-\cos\theta}{1-\cos\theta_f} \times
      \frac{1+\cos\theta_f}{1+\cos\theta} \right)\right)^{1/4} ~\keV, \\
\end{array}
\end{equation}
\noindent
where 
$$
\begin{array}{l}
f_\alpha = \left( 3 (1+\alpha) \right)^{1/2(1+\alpha)}
(2+\alpha)^{((2+\alpha)(\gamma-1)-1)/4} \times \\
\qquad{} \qquad{} \times \left(
\Omega^\prime\right)^{-(3+\alpha)/4(1+\alpha)}.\\
\end{array}
$$

Characteristic temperatures derived above are fairly consistent with
those reported by \citet{poutanen}.
For a reasonably high $\mdot =10^3$ and $\theta_f = 0.4\,\rad$ the outer photosphere
temperature is $\sim 10^5\,\rm K$. Together with a photospheric
luminosity $\sim 10^{39}\ergl$ this makes the wind photosphere both a
bright UV/optical and a bright extreme ultroviolet (EUV) source capable for ionizing
large amounts of interstellar and circumstellar gas. We discuss the
properties of HII-regions created by supercritical accretor 
wind photospheres in section~\ref{sec:nublado}.
Very shallow temperature decline makes the spectrum flat from $E \sim
0.005~\keV$ to $\sim 1~\keV$.
The highest temperature value is predicted by accretion disc theory and is much
higher than the temperature at spherisation radius \citep{poutanen}:

\begin{equation}\label{E:maxT}
T_{max} = 1.27 M_{10}^{-1/4}~\keV .
\end{equation}
\noindent


\section{Effects of Irradiation}\label{sec:irrad}
 
In the standard disc model irradiation is a minor, often negligible effect as
the disc is thin and the fraction of emitted luminosity absorbed by the outer
parts of the disc is of order of $O(H/R)^2$. On the contrary, in the
case of supercritical funnel most of the radiation emitted at any point on the
wall surface will be absorbed again or reflected. The probability of re-absorption is of the
order  $\cos \theta_f$, hence the energy balance will be significantly
affected by irradiation.
As long as true absorption dominates
over electron scattering, $\sigma \gg \sigma_T$, all the quanta may be
considered absorbed and re-emitted by the walls remaining a locally
blackbody source.



The energy input due to incident radiation is characterized by the flux
$F^\prime (R)$ normal to the wall surface and directed inward, which may
originate both from the funnel bottom and inner parts of the walls.
In general, the incident flux may be expressed as
\begin{equation}\label{E:def:frel}
  F^\prime = F \int \left( \frac{T_{eff}(\mathbf{R^\prime})}{T_{eff}(\mathbf{R})}
  \right)^4 \frac{ \left| (\mathbf{n \cdot d}) (\mathbf{n^\prime \cdot d}) \right|}{d^2}  dS^\prime
\end{equation}
where the meaning of $\mathbf{n}$ and $\mathbf{d}$ vectors is illustrated in
figure~\ref{fig:irradscheme} for both wall and bottom irradiation.


\subsection{Problem Formulation}

In this section Cartesian and spherical coordinates are used. 
$z$ axis coincides with the symmetry axis of the disc, funnel and jets. 
The remaining degree of freedom is adjusted to set to zero the
azimuthal coordinate of the point under consideration, where the
incident flux is calculated. 
This point is set by a radius vector $\mathbf{R}$,

$$
\mathbf{R}=R \left(\begin{array}{c}\sin \theta \\ 0 \\ \cos \theta \\\end{array}\right)
$$

In the same way in the following subsections we define the bottom
radius vector $\mathbf{R_0}$ and the radius vector $\mathbf{R^\prime}$
of the variable point on the funnel walls.

$$
\mathbf{R_0}=R_0 \left(\begin{array}{c}0 \\ 0 \\ \cos \theta
  \\\end{array}\right)
\qquad{}  
\mathbf{R^\prime}=R^\prime \left(\begin{array}{c}\sin \theta \cos \varphi \\ \sin \theta \sin \varphi \\ \cos \theta \\\end{array}\right)
$$

The geometry of funnel self-irradiation is shown in
figure~\ref{fig:irradscheme}.
Variable infinitesimal plate with area $dS^\prime$, radius vector
$\mathbf{R^\prime}$ and normal $\mathbf{n^\prime}$ contributes to the
incident flux in the current point by:

$$
dF^\prime = \left| (\mathbf{n \cdot d}) (\mathbf{n^\prime \cdot d})
\right| \frac{dS^\prime}{2 \pi d^4},
$$
\noindent
where $\mathbf{d = R^\prime - R}$, $d = |\mathbf{d}|$.
Due to surface brightness invariance from distance it is convinient to
use dulition factor $w$ depending on geometry only and to normalize
the incident flux over the outcoming flux $F(\mathbf{R})$ in the
current point. 

$$ 
w =  \frac{ \left| (\mathbf{n \cdot d}) (\mathbf{n^\prime \cdot d}) \right|}{d^2}
$$ 

Incident flux in the units of the outcoming integral flux ($f =
F^\prime / F$) can be expressed as an integral over the funnel inner surface:

\begin{equation}\label{E:def:frelwalls}
f = \int \left( \frac{T_{eff}(\mathbf{R^\prime})}{T_{eff}(\mathbf{R})}
\right)^4 \frac{ \left| (\mathbf{n \cdot d}) (\mathbf{n^\prime \cdot d}) \right|}{d^2}  dS^\prime
\end{equation}

In the case of the funnel bottom we avoid integration suggesting the
bottom a point-like source emitting as a flat plate. 


\subsection{Irradiation by the Funnel Bottom}

\begin{figure*}
\centering
\FigureFile(150mm,60mm){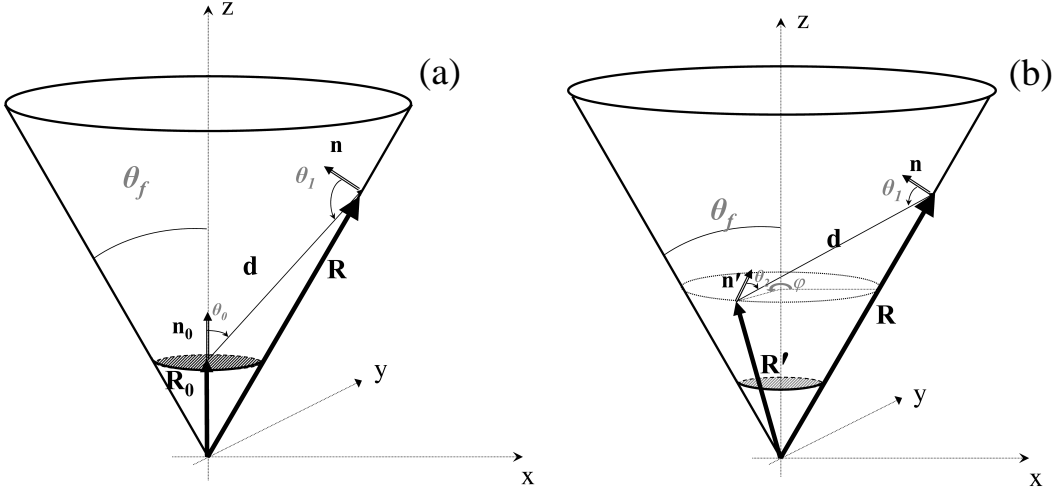}
\caption{
Schemes explaining the geometry of irradiation effects. (a)
irradiation by the bottom. (b) irradiation by the funnel walls.\label{fig:irradscheme}
}
\end{figure*}

The first heating term is simpler to handle because the source is
practically point-like and its temperature is not affected by the
funnel walls. Due to this, one can do without integration. 
Funnel bottom can be considered a spherical surface with
the surface area  $S_{sph}= 2 \pi (1-\cos \theta) R_0^2 $ or a flat
circular plate of radius $R_0 \sin\theta$ (because $R_0$ is the
starting radial coordinate along the wall, not along the $z$-axis) with
the surface area  $S= \pi R_0^2 \sin^2\theta$.
Both expressions give similar results when $\theta \ll 1$.
Here we assume the bottom having flat surface (see figure~\ref{fig:irradscheme}a).
The photon source coordinate and normal unit vector are:

$$
\mathbf{R_0}=R_0 \left(\begin{array}{c}0 \\ 0 \\ \cos \theta \\\end{array}\right)
\qquad{} \qquad{}
\mathbf{n_0}= \left(\begin{array}{c}0 \\ 0 \\ 1 \\\end{array}\right)
$$
\noindent

Distance vector connecting emitting and receiving
  points:

$$
\mathbf{d} = \mathbf{R-R_0} = R_0 \left(\begin{array}{c}  r \sin \theta
  \\ 0 \\ (r-1) \cos \theta \\\end{array}\right)
$$
\noindent
Here $r=R/R_0$. Dilution factor in that case:

\begin{equation}\label{E:dilubot}
\begin{array}{l}
w(x,\theta, \varphi)=\frac{|(\mathbf{n}\cdot\mathbf{d})(\mathbf{n_0}\cdot\mathbf{d})| S}{ 2 \pi d^4}=\\
\qquad{} =\frac{(r-1) \sin\theta \cos^2\theta}{2 \left(1+r^2-2r\cos\theta \right)^{2}}
\end{array}
\end{equation}
\noindent

Total normalized irradiating flux from the bottom can be therefore expressed as:

\begin{equation}\label{E:botflux}
 f_{bottom} = \frac{1}{2} \frac{(r-1) \sin^3\theta
 \cos^2\theta}{\left(1+r^2-2r\cos\theta \right)^{2}} \tau^{-4}
\end{equation}
\noindent

\subsection{Self-Irradiation by the Funnel Walls}

The energy input from the absorbed photons is directly calculated
using equation (\ref{E:def:frel}). Surface element is an elementary area on
a conical surface, $dS^\prime = 2\pi \sin\theta R^\prime dR^\prime
d\varphi$. Radius vectors and normals are defined. Distance vector:

$$
\mathbf{d} = \mathbf{R^\prime} - \mathbf{R} = R \left(\begin{array}{c}
   \sin \theta (1-x\cos\varphi)
  \\ -x\sin\theta \sin\varphi \\  \cos \theta (1-x) \\\end{array}\right)
$$

\noindent
where $x=R^\prime / R$. Dilution factor in this cases takes the form:

\begin{equation}\label{E:diluwalls}
\begin{array}{l}
w(x,\theta, \varphi)=\frac{|(\mathbf{n}\cdot\mathbf{d})(\mathbf{n^\prime}\cdot\mathbf{d})| S}{ 2 \pi d^4}=\\
\frac{1}{2\pi R^2}
\frac{x \sin^2\theta \cos^2 \theta (1-\cos\varphi)^2}{2 \left( (1-x)^2
  + 2x (1-\cos\varphi) \sin^2 \theta\right)^{2}}\\
\end{array}
\end{equation}

Absorbed flux is given by an integral over $x$:

$$
F^\prime (R) = 2\pi \sin\theta \int_{R_{in}/R}^{R_{out}/R} F(R x) I(x,\theta) x dx 
$$

\noindent
where $I = \int_{-\pi}^{\pi} w(x,\theta,\varphi)
d\varphi$. Integration over $\varphi$ is straightforward however
complicated, so we leave it for appendix \ref{sec:app:phint}. 
Finally, normalized absorbed flux can be expressed as:

\begin{equation}\label{E:walint}
\begin{array}{l}
f_{walls} (R) = \frac{\cos^2 \theta}{4 \sin\theta} \int_{R_{in}/R}^{R_{out}/R}  
\left( \frac{T_{eff}(\mathbf{R^\prime})}{T_{eff}(\mathbf{R})}
\right)^4 \times \\
\qquad{} \times \left( 1 - \frac{|1-x|}{\sqrt{1-2x \cos(2\theta) +x^2}} \frac{1-2x (1-3 \sin^2 \theta) +x^2}{1-2x \cos(2\theta) +x^2} \right) dx\\
\end{array}
\end{equation}

\begin{figure}
\center{
\FigureFile(\columnwidth,\texthight){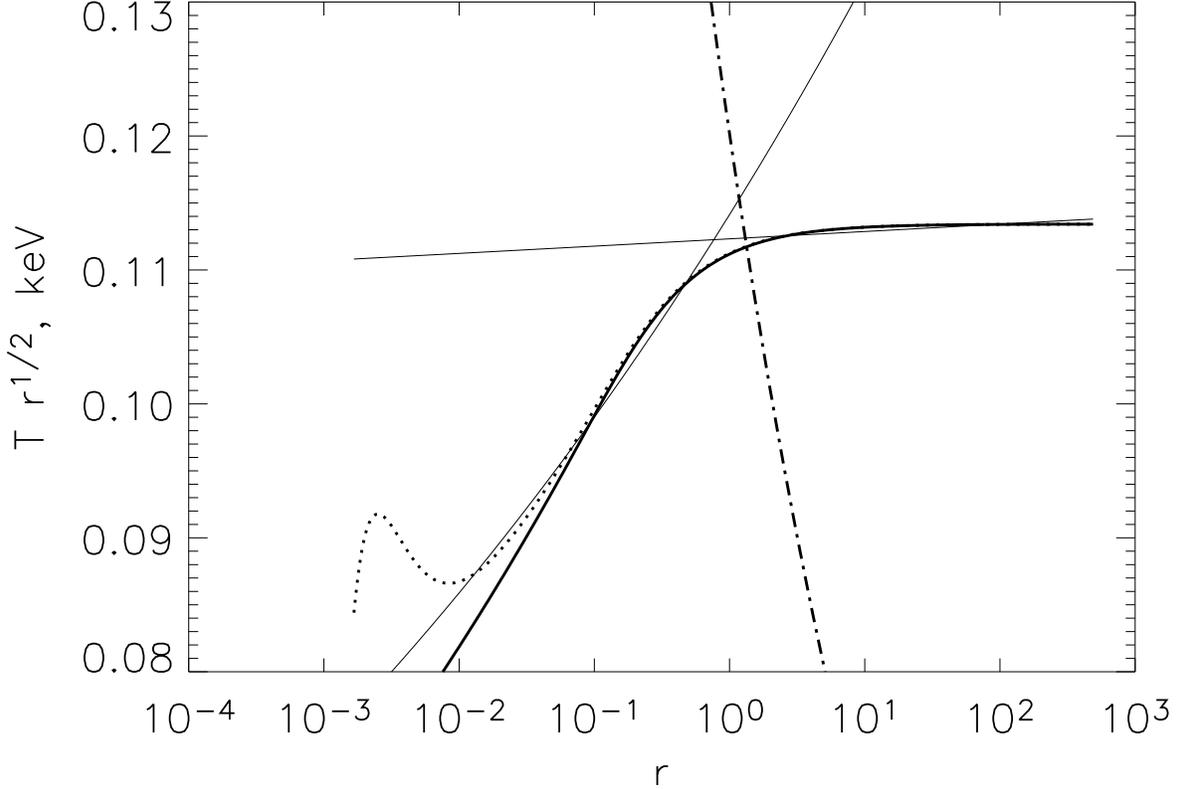}
\caption{
Funnel wall temperature dependence on radius, normalized by $r^{-1/2}$. Thick solid line: without
irradiation, thick dotted line: with irradiation effects included,
three iterations. Thin solid
lines represent the best power-law fits to the irradiated funnel wall temperature at larger
$r$ ($>1$) and lower $r$ ($10^{-2} < r < 1$), $T \propto r^{-0.44}$ and $T \propto r^{-0.50}$, correspondingly. Dot-dashed line represents the
$T\propto r^{-3/4}$ temperature slope characteristic for standard
discs.
\label{fig:tempir}
}
}
\end{figure}

\begin{figure}
\center{
\FigureFile(\columnwidth,\texthight){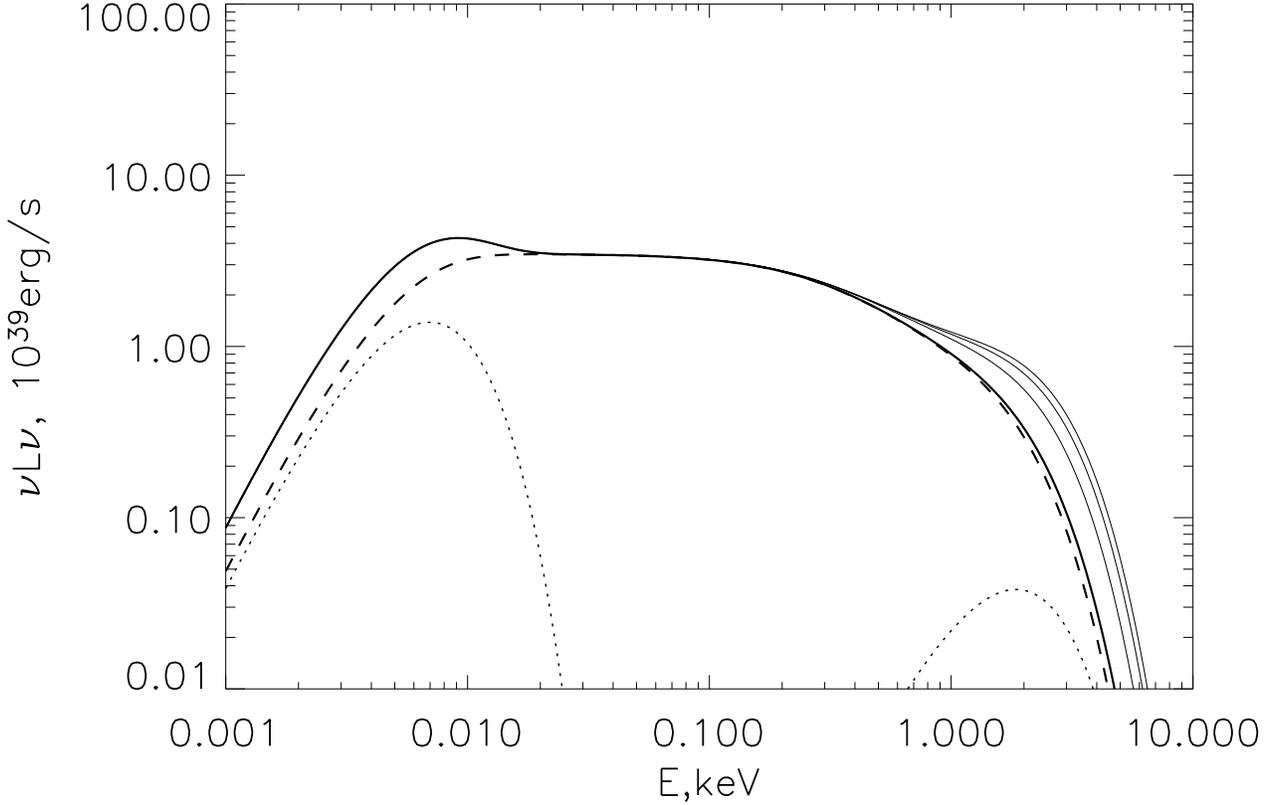}
\caption{
Irradiation effects on the outcoming spectrum. Solid lines show the
spectral energy distribution of a face-on supercritical funnel for
different number of iterations (0, 1, 2 and 3). Funnel
walls without irradiation give the spectrum shown by dashed lines. Dotted lines correspond
to the contributions from the bottom (at high energies) and the outer photosphere.
Parameters are the same as in figure \ref{fig:2d}. \label{fig:spectra:irrad}
}
}
\end{figure}

Irradiation is treated iteratively. Three to five iterations generally
sufficient for convergence with accuracy 
better than 5\%. As may be seen in figures
\ref{fig:tempir} and \ref{fig:spectra:irrad}, mostly the inner parts
of the funnel walls (and subsequently mostly the X-ray range) 
are affected. Incident flux produced at large radii by the irradiating
funnel bottom and inner parts of funnel walls changes as $F^\prime \propto
r^{-3}$ (see equation (\ref{E:botflux})), while the predicted
temperature dependence on radius implies $F^\prime \propto
r^{-2}$  for the emergent flux, therefore the effect is expected. 


\section{Emergent Spectra and Spectral Variability}\label{sec:sp}

Here we calculate the outcoming spectra as functions of inclination
and supercritical wind parameters. All the surfaces are considered
locally blackbody sources and the observed spectrum (in terms of
apparent isotropic luminosity $\nu L_\nu$) is calculated by
integrating over all the visible surfaces applying visibility
conditions. The most important effect is that at inclinations $i
\gtrsim \theta_f$ funnel bottom and some part of the funnel wall
become obscured from the observer.

Doppler boosts are accounted for approximately. We consider funnel walls
moving at an angle $\theta_f$ with the line of sight (that is exactly
true for a face-on object), and the material consisting the 
funnel bottom moving at an angle $i$. The velocity is determined
according to equation (\ref{E:velty}) and is considered equal for
the funnel bottom and the wind at $r_{in}$. 
Photon energies are modified by a factor:

\begin{equation}\label{E:delta}
\delta = \frac{1}{\gamma (1-\beta \cos\theta)}
\end{equation}
\noindent
where $\beta =  v/c$ is the dimensionless velocity, $\gamma =
1/\sqrt{1-(v/c)^2}$ is the Lorentz factor, and $\theta$ is $\theta_f$
for the walls and $i$ for the bottom. 
Photon numbers are modified by a factor $\delta$, and the time
compression factor is not relevant here because the flow is
stationary. Apparent luminosity $L_E$ changes proportionally to
$L_E(E) \propto \delta^2 L_E^{(o)}(E/\delta)$, where $L_E^{(o)}$ is
the apparent luminosity calculated without relativistic effects.
 Note that Doppler boosts
become important only when the velocities in the inner parts
of the flow are relativistic (either when $\mdot \lesssim 10$ or for
$\alpha \lesssim -0.5$). See also discussion in section
\ref{sec:relat}. 

In figures \ref{fig:tempir} and \ref{fig:spectra:irrad} we compare the
temperature profiles and SEDs with and
without irradiation. We adopt $\alpha=0$, $\mdot=10^2$, $r_{in} =
1 / 6\mdot = 1.7 \times 10^{-3}$, $\theta_f = 23^\circ$.
The temperature dependence on radius is close to
a broken power-law in both cases, and the net effect of irradiation is
in altering the mean temperature in the innermost parts of the funnel
by about 20\%.
In figure  \ref{fig:spectra:irrad} SEDs are calculated for zero
inclination and different number of iterations used to account for
irradiation effects.
Evidently, only the
funnel walls is affected by irradiation that alters the flux by a
factor of $2\div 3$. Spectrum is practically flat in the EUV/soft
X-ray region but curves near the $T_{sph} \sim 0.1\,\keV$. High-energy
cut-off is present at several \keV\ if the inner radius is $r_{in} \sim 1/\mdot$.

\subsection{Self-Occultation}\label{sec:incl}\label{sec:obscur}


Let $i$ be the inclination of the funnel (angle between the symmetry
axis and the line of sight). 
For any given radial coordinate at the funnel wall surface three
cases are possible: the annulus is fully visible (this is
possible only for $i < \theta_f$), invisible or partly visible.
If the inclination is larger than $\theta_f$ for different radii
either second or third case takes place for every annulus. 
Visibility may be quantitatively described by a factor $y$ defined as
the visible part of the given annulus (having constant distance $r$ from
the center). It may be determined by an analytical ray-tracing method.

$$
\vector{r} = \vector{R} + s \vector{l}
$$
\noindent
Here $\vector{l}$ is a unit vector directed towards the observer (we
use the same scheme as in the previous section), $\vector{R}$ is the
radius vector of the intersection point between the funnel surface and
the ray starting from $\vector{r}$ and directed
towards the observer. Boundary case when the intersection point has
the radial coordinate equal to the photosphere radius is of interest
here. Normalizing over the outer radius ($x = r / r_{out}$), one can
express the vectors as follows. 

$$
\vector{r} = x \left(\begin{array}{c}
\sin\theta_f \cos A \\ \sin\theta_f \sin A \\ \cos \theta_f
\end{array}\right)
$$
$$
\vector{R} =  \left(\begin{array}{c}
\sin\theta_f \cos A_0 \\ \sin\theta_f \sin A_0 \\ \cos \theta_f\\
\end{array}\right)
$$
$$
\vector{l} =  \left(\begin{array}{c}
\sin i \\ 0 \\ \cos i\\
\end{array}\right)
$$
\noindent

It is easy to express two of the unknowns ($s$ and $A_0$) from the
others and obtain the solution for $A$ (azimuthal coordinate of the
starting point):

$$
\sin A = \frac{1}{2} \left( \frac{\tan \theta_f}{\tan i}+\frac{\tan
  i}{\tan \theta_f}\right) +\frac{1}{2x} \left( \frac{\tan \theta_f}{\tan i}-\frac{\tan
  i}{\tan \theta_f}\right)
$$
\noindent

If the annulus is partially visible $y$ may be calculated as the
length of the arc (devided by $2\pi$) connecting the two points
defined by the solutions of the equation above. 
Finally:

\begin{equation}\label{eq:obscur:weights}
y = \left\{
\begin{array}{cr}
0 & K<-1\\
\frac{1}{2}+\frac{1}{\pi} \arcsin K & -1 \ge K < 1\\
1 & K \ge 1\\
\end{array}
\right. ,
\end{equation}

\noindent
where:

$$
K = \frac{1}{2} \left( \frac{\tan \theta_f}{\tan i}+\frac{\tan
  i}{\tan \theta_f}\right) +\frac{1}{2x} \left( \frac{\tan \theta_f}{\tan i}-\frac{\tan
  i}{\tan \theta_f}\right)
$$

\begin{figure}
\center{
\FigureFile(\columnwidth,\texthight){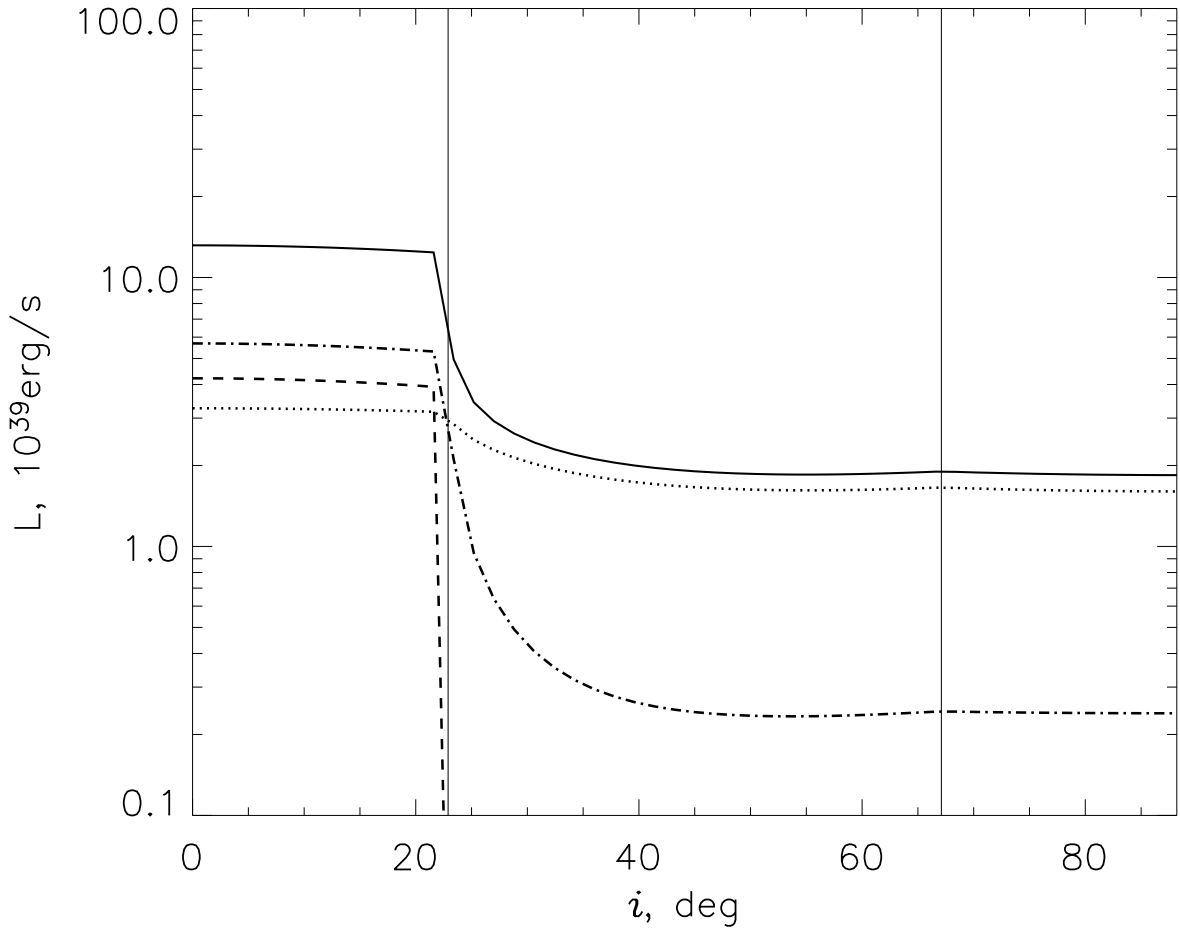}
\caption{
Apparent isotropic luminosity as a function of inclination. Solid line
represents bolometric luminosity. Dashed, dot-dashed and dotted lines
represent harder X-rays ($0.4\div 20~\keV$), softer X-rays ($0.1\div
0.4~\keV$) and EUV ($0.01\div 0.1~\keV$). Vertical lines mark
$\theta_f$ and $\pi/2-\theta_f$.\label{fig:incl}
}
}
\end{figure}

\begin{figure}
\center{
\FigureFile(\columnwidth,\texthight){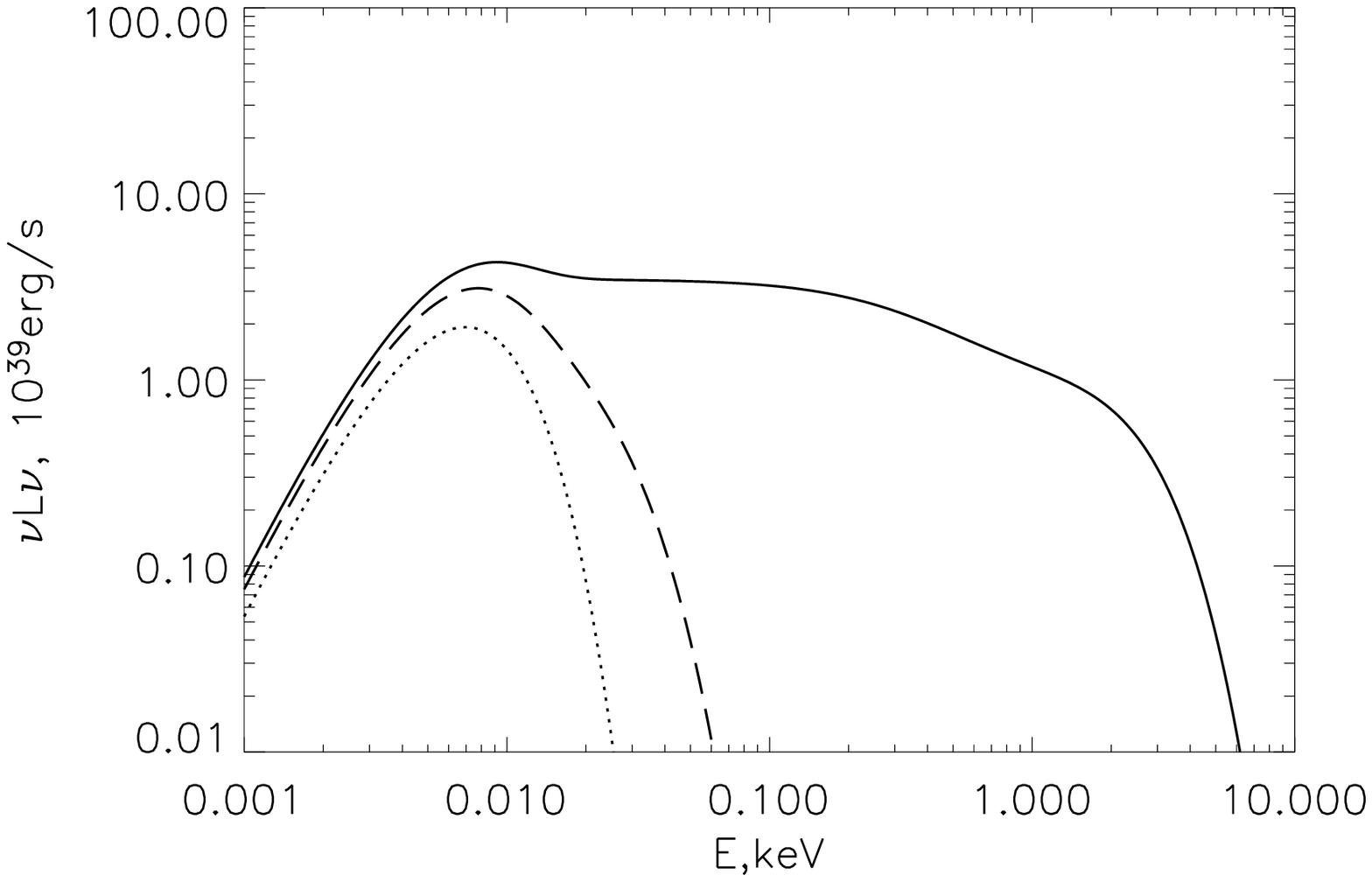}
\caption{
Inclination effects on the outcoming spectra.
Apparent values of $\nu L_\nu$ are shown for $i = 0$, $25$, and
$90$ degrees as a solid,
dashed and dotted lines, respectively. 
\label{fig:spectra:incl}
}
}
\end{figure}

\subsection{Implications of Self-Occultation Effects}

\citet{wata} studied self-occultation effects for fat
accretion discs coming to an evident conclusion that the spectrum
becomes softer at large inclinations. \citet{heinz} comes to a similar
conclusion considering the results of 2D radiative hydrodynamic simulations by
\citet{ohsuga2005}. 
In our case this effect is even more profound (see figure~\ref{fig:incl}): at inclinations $i \sim
\theta_f$ the X-ray component abruptly disappears (because only the
inner parts of the funnel contribute significantly to the X-ray range) and only the EUV
component may be observed.

Due to that reason some supercritical accretors (those viewed at large inclinations)
will not show the ULX phenomenon (see figure~\ref{fig:spectra:incl})
but remain luminous UV sources.
For $\theta_f=0.4\,\rad$ the number of orphan ultraviolet sources of
similar nature is about $\cos
\theta_f / (1 - \cos\theta_f) \sim 10$ times higher than the number of
ULXs. 

Observations of these sources in the EUV are complicated by neutral
gas absorption. 
In \citet{mf16} we discuss the possibility of far-ultraviolet
observations with \galex\ \citep{martingalex} and \hst\ in the ultraviolet.
Supercritical accretors prove to be difficult but still reachable
targets for  \galex\ and \hst, appropriate for pointing observations.

We do not know yet how much there is similarity between ULXs and
SS433. 
SS433 
exhibits jet/disc
axis direction variations with a super-orbital
precession period about $160\,^{\rm d}$. 
Similar effects are observed for some other X-ray binaries
\citep{superorb}.
There are indications that the supercritical disc (and consequently
the wind with the funnel) follows the motions of the jets.

\begin{figure}
\center{
\FigureFile(0.8\columnwidth,\texthight){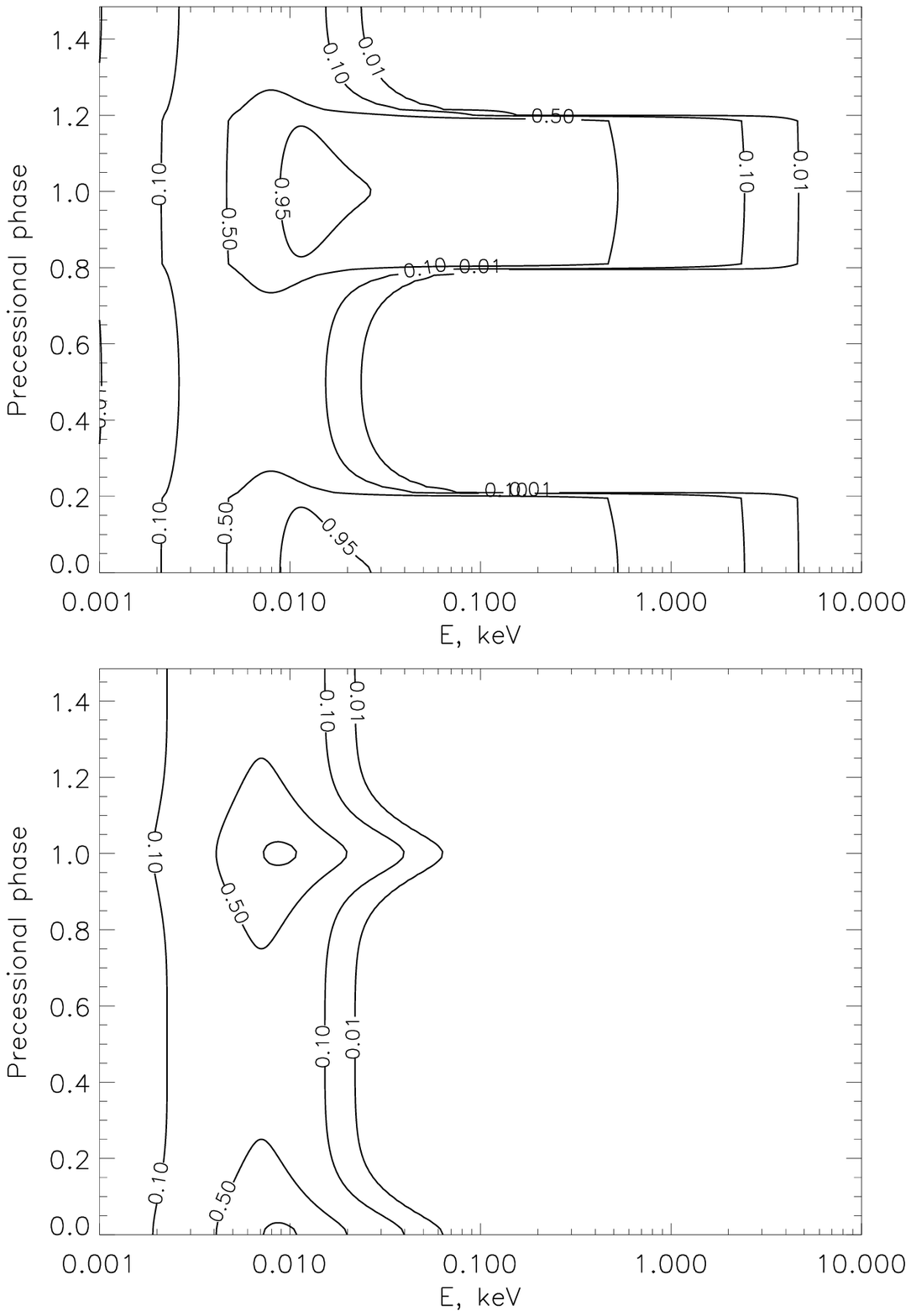}
\caption{
Spectral variability of precessing supercritical accretion disc
funnel. Lines of constant flux ($EF_E$)
as a function of energy and precessional phase are given. Flux is
normalized over its maximal value. $\theta_f=22.9\deg$, $\mdot=10^3$,
$\alpha=0$. Upper diagram corresponds to $i_0=20\deg$ and
$i_A=20\deg$, lower has $i_0=45\deg$ and $i_A=20\deg$. In the latter
case funnel bottom is never visible and funnel walls always remain at least
partially occulted. \label{fig:prec}
}
}
\end{figure}

A nearly face-on supercritical accretion disc may show periodical 
X-ray variability by one-two
orders of magnitude when $\theta_f$ is less than the maximal and
higher than the minimal value of $i$. 
Precessional variability should have a distinguished light curve with
a flat maximum and continuous flux change near the minimum light. In
figure~\ref{fig:prec} we present the predicted X-ray spectra as
functions of precessional phase for two sets of parameters. We assume
here a simplified version of the kinematical model applied to SS433
\citep{ss_cinema}:

$$
\cos i = \cos i_A \cos i_0 + \sin i_A \sin i_0
\cos (2\pi \psi),
$$
\noindent
where $i_0$ is the inclination with respect to the precession axis, $i_A$ is the
amplitude of inclination variation, and $\psi$ is the phase of
super-orbital period.

The only ULX exhibiting variability that may be considered
super-orbital is X41.4+60 in M82 \citep{m82_kaa}. {\it RXTE} X-ray
flux varies by $\sim 50$\% for this source  with a $62^{\rm d}$ period. 
Irregular variability on similar timescales is much more ordinary
among ULXs \citep{hoix_parola} possibly indicating that the funnel shape itself changes
rapidly enough to level the effects of precessional variability.


\section{Comparison with Observations}\label{sec:obs}



\subsection{Data on ULXs}\label{sec:data}

\xmm\ datasets on two X-ray sources, NGC4559~X-7 and NGC6946~ULX-1 were
analysed. 
We used archival \xmm\ EPIC \citep{epic} data (MOS1,2 and PN detectors) 
obtained on 8 October 2006
(observation ID~0152170501) and 25 June 2004 (observation
ID~0200670401), correspondingly.
All the data were reduced using standard XMM-Newton Science Analysis
System (SAS) procedures. 
Response files were made using SAS tools {\tt rmfgen}  and {\tt arfgen}. 
We set \textit{flag}$=0$ for PN data. We used \textit{pattern} $\le 4$ for PN and $\le
12$ for MOS data. 

NGC4559~X-7 is known as a ``supersoft'' bright ULX \citep{ngc4559_sp}. 
It does not coincide with a bright stellar or nebular optical
counterpart. However, there are indications that the source is
connected with $\sim 10\,\Myr$ old stellar population \citep{ngc4559x7_soria}.

NGC6946~ULX1 \citep{ngc6946_RoCo} is known to be a source of
moderate luminosity $L_X \sim 3\times 10^{39}\ergl$ with
an X-ray spectrum usual for ULXs. 
Stellar optical counterpart was detected by {\it HST} in the visible
range \citep{BFS} as a relatively bright star ($M_V \lesssim -7^m$). 
As many other ULXs, NGC6946~ULX1 coincides with a bright optical
nebula. \citet{mf16} analysing the emission spectrum of the nebula
conclude that the central source may be a hard EUV source with
temperature $T \sim 10^5 \rm K$ and luminosity $L \sim 10^{40}$\ergl,
roughly consistent with what one may expect from a $\mdot \sim
10^2\div 10^3$ supercritical accretor.  

Three models were used to fit the data: standard disc \citep{discbb} +
power law, p-free disc \citep{discpbb} and self-irradiated multi-color funnel model
presented here. Efficiency of the p-free model for ULXs was already shown by
\citet{vier2006} and other works. Because the temperature of the funnel walls decays in a
practically power-law manner (with $p \sim 0.5$, see figure
\ref{fig:tempir}) one should expect p-free model to produce SEDs close to
the SEDs predicted by our model. However, the temperature distribution required is very flat,
so we have shifted the lowest allowed $p$ from $0.5$ to $0.05$. 
For each object all the three extracted spectra were fitted
simultaneously. Spectral ranges $0.1-12$\,\keV\ for PN and
$0.1-10$\,\keV\ for MOS data were used. Photoelectric absorption by a
solar metallicity material was included as a free parameter in all the
models. 

For the supercritical funnel model used for X-ray data fitting we
fixed the velocity exponent ($\alpha = -0.5$), the outer radius
($r_{out} = 100$) and mass ejection rate (to $\mdot = 10^4$) 
and the inclination to $0^\circ$. 
For fixed $T_{in}$, $r_{in}$ and $r_{out}$ values the only effect of
the mass accretion rate is in changing the gas velocities. As long as
$v(r_{in}) \ll c$, the mass ejection/accretion rate 
\mdot\ does not affect the shape of the X-ray spectrum. 

Low values of $\alpha \sim 0 .. -0.5$ 
provide better fits to the data. However, we fix the parameter to
avoid degeneracy with $T_{in}$ and $r_{in}$.
$\alpha = -0.5$ is expected if the velocity is
proportional to the virial velocity at any given radius. In ballistic
approximation (when the particles of the wind are first rapidly
accelerated and then move along hyperbolic trajectories in the
gravitational field of the accretor)
$\alpha \simeq -0.5$ in the inner parts of the flow and approaches 0 at
larger radii.  


In table \ref{tab:xspectra} we present the results of spectral fitting. 
It may be seen that in both cases a two-component standard
disc + power law model gives the best results because of being
capable to fit the harder part of the spectrum that is difficult to
handle using only thermal models with exponential high-energy
cut-offs. 

Because the best-fit temperature in both cases is rather high
$T_{in} \sim 1\div 2\,\keV$, we expect the inner parts of the funnel
to be practically transparent and $R_{in}$ ti be close to the last stable
orbit radius. In this assumption, $r_{in}$ may be directly converted
to the mass accretion rate.
If we equate the inner radius to the last stable orbit for a
Schwarzschild black hole,

$$
\mdot = \frac{\cos\theta_f}{18} \frac{R_{sph}}{R_G} \simeq \frac{1}{6\,r_{in}} 
$$

The estimated dimensionless mass ejection rates are therefore about
100 for NGC4559~X-7 and about 300 for NGC6946~ULX-1,
correspondingly. The latter value is consistent by the order of
magnitude with the mass accretion rate estimate for NGC6946~ULX-1
resulting from the optical spectroscopy of the nebula MF16 associated
with this X-ray source \citep{mf16}. 


The well-known high-energy curvature \citep{curva} can be explained by
$T_{in} \sim 1\div 2.5\,\keV$. If a hardening factor $\sim 1\div 2$ is used
(see also section \ref{sec:trep}),
the inner temperature may be similar to the expected maximal
temperature for a critical accretion flow around a $\sim$10\Msun\ black hole
given by equation (\ref{E:maxT}) or even slightly lower.
High inner temperatures like $10\,\keV$ also result in acceptable fits.
Model is insensitive to $r_{out}$ save for the cases when this
parameter is $\lesssim 1$ therefore we fixed the parameter to $100$. 


Low values of $r_{in}$ appear to be a real feature of ULXs. Funnel interior
is expected to be practically transparent to the X-rays down to several
gravitational radii, having
rather deep-lying bottom pseudo-photosphere (or none at all). Observations in a broader spectral range
are needed to distinguish between the thermal radiation from the immediate
vicinity of the compact object and comptonisation effects. 


Though the model fits the data in a quite acceptable way its
parameters are poorly constraint. The spectral shape does not depend
very much on the actual accretion rate and velocity law.
There are numerous ways of making the spectrum harder: taking into
account the difference between mass accretion and ejection rates,
applying comptonisation effects etc.


\section{Discussion}\label{sec:disc}\label{sec:trep}

\subsection{Limitations of the Model}

Our calculations do not account for
radiation feedback on the dynamics. That is not only wind
acceleration within the regions of interest but also evaporation from the
funnel walls and development of instabilities in the outer parts of
the funnel in the strong radiation field of the inner parts. The
instability may resemble that of accretion discs with irradiation (see
\citet{min93} and references therein) but the wind is
unable to influence the irradiating source and its role in developing
the instability is purely passive. 


The structure of the inner parts of the flow may be much more
complicated than we assumed. At $r \sim 1$ the source of the wind
becomes spatially resolved and our radial approximation fails. Besides
this, the angular distribution of the energy influx may be more
complicated and higher angular harmonics may appear.
 
Qualitatively  the effects of non-zero size mass loading region may be
considered as follows: let us assume that $\mdot \propto r$. The main
radial equation for $k=0$ takes the form:

$$
  \partial_r u - \frac{1}{r^{2+\alpha}}
  \partial_r \left( r^{3+\alpha} \partial_r u\right)
  + \frac{\gamma (2+\alpha) }{r} u =0
$$
\noindent
It is easy to check that the equation allows two power-law solutions
($u\propto r^\sigma$, where $\sigma = - 0.5 \left( (1+\alpha) \pm
\sqrt{(1+\alpha)^2 + 4 \gamma (2+\alpha)}\right)$) corresponding to
inward and outward diffusion. For $r\ll 1$ and $\alpha=-0.5$ energy density decreases as
$u \propto r^{-1.68}$ instead of $u \propto r^{-2}$. The difference
increases for higher $\alpha$. 

We do not account for the emission of the hot interior of the funnel
as well as for thermal comptonisation effects in the wind and pure
reflection. All these effects are likely to harden the outcoming
spectra making inner temperature estimates from the observational data
shifted towards higher values. Hardening factor $T_h \simeq 2.6$ measured for
standard discs \citep{bor99} together with relativistic effects 
may lead to the inner temperatures about 1\,\keV\ appear as several \keV. 

Reflection and absorption by moving partially ionized gas may simulate the
soft-excess observed in many ULX spectra \citep{soft_excess} in a way it
was proposed by \citet{done} for AGNi and recently by
\citet{goncalves} for ULXs. 
Understanding the structure of the outer photosphere of the wind
requires more complicated modelling taking into account both
significant non-sphericity of the outflow and various opacity
sources. 


\subsection{Relativistic Effects}\label{sec:relat}

Broadly speaking, relativistic effects are expected to change photon
energies and fluxes by factors close to $\delta = \left(\gamma
(1-\beta \cos \theta_f) \right)^{-1}$. Here, $\beta$ and $\gamma$
correspond to the local dimensionless velocity and Lorentz-factor. 
The temperature of the X-ray spectrum is additionally
increased by:

\begin{equation}
\frac{\Delta T}{T} = \delta - 1 \simeq 5\times 10^{-3} \cos^{3/2} \theta_f
\mdot_3^{-1/2} r_{in}^{\alpha}
\end{equation}

The effect becomes significant if the velocity is relativistic in the
inner parts of the flow (that corresponds to $\alpha \lesssim -0.5$ in
the above formula). The total observed luminosity 
is altered by a factor $\delta^3$ ($\delta^2$ factor from geometrical
reasons and one $\delta$ from the increased energy of the
photons). For $\mdot = 10^3$ the effect is about one percent if the
velocity at the spherisation radius is considered. However, if the inner
parts of the wind have mildly relativistic velocities, the
X-ray part of the spectrum appears about $2\div 3$ times brighter
for $\beta \sim 0.3$, $\theta_f \sim 20^\circ$.

The funnel bottom is the most likely part of the flow to be affected
by relativistic effects. The gas inside the funnel cone is supposed to
move with mildly relativistic velocities \citep{EGK}, approximately towards the
observer at low inclination angles. The observed colour temperature
changes by a factor of $\delta \simeq 1+\beta \simeq 1.1 \div 1.5$. 
The total observed luminosity becomes several times higher. 
Our model allows to include the relativistic Doppler effect in calculations accurately
for $i=0$ (see section \ref{sec:sp}). At non-zero inclinations the
deviations from our approximation (that all the funnel wall material
moves at $\theta_f$ angle with respect to the line of sight) are of
the order $O(\beta \sin i)^2$.

Relativistic aberration is also able to dump irradiation
effects. The X-ray part of the spectrum will be the most
affected. Irradiation becomes considerably smaller if the radiation
emitted by the moving gas becomes anisotropic enough. However, for
mildly relativistic flows ($\beta \lesssim 0.5$) the effect is not so
severe compared to the effects mentioned above. For example, the 
irradiating flux from the funnel bottom changes for $r\simeq r_{in}
\simeq 1/6\mdot$ and $\alpha = -0.5$
by about $\gamma^3 - 1 \simeq 20\div 30\%$. At large radii the effect is
smaller and may change sign at certain radii.

\subsection{Observational Predictions for ULXs}

There are several effects expected if our estimates are
correct:
{\it (i)} supercritical accretors viewed at low inclinations
  ($i \lesssim \theta_f $) will be seen as ULXs ($L \gtrsim 10^{39}$\ergl);
{\it (ii)} independently of its inclination a supercritical accretor is
  a luminous UV source;
{\it (iii)} predicted X-ray spectra of supecritical accretors viewed
  face-on are similar to p-free model spectra with $p \sim 0.4\div0.6$;
{\it (iv)} because of high EUV luminosities supercritical accretors
  should ionize the wind above the photosphere and 
  establish Str\"omgren zones;
{\it (v)} if the accretion disc precession characterising SS433 is
  usual for supercritical accretors viewed nearly face-on, strong X-ray flux modulation is
  expected for at least a number of sources with ``super-orbital''
  periods like tens and hundreds of days.

Most of the observational properties of ULXs are naturally explained
in our model. There is observational evidence that at least some ULX nebulae
 are powered by photoionisation from the central source having
luminosity comparable with the apparent X-ray
luminosity. IMBH binaries should have difficulties in ionizing the
surrounding gas unless the mass of the IMBH is very high, $\gtrsim
10^4\Msun$ (see discusion in \citet{mf16}).

For very high accretion rates $\mdot \gtrsim 10^3$ the outer
photosphere is very large, comparable to the probable size of the
binary system. 
Applying $\alpha=0$ and $\theta_f=0.4\,\rad$ results in physical radius
values:

\begin{equation}
R_{out} \simeq 4 \times 10^{13} \mdot^{3/2}_3 \rm cm .
\end{equation}

However in a real high-mass binary conditions the structure of the
outflow is preturbed by tidal forces and
the outflowing gas itself rapidly recombines (see next section).
Because the flow is essentially supersonic and perturbed by a strong
non-axisymmetric potential it is likely to become highly
inhomogeneous. The equatorial outflow of SS433 is an example of such
kind. 
The photosphere size in the optical then saturates at a radius of the
order $10^{12}\rm cm$  preventing wind photospheres
from becoming ``red hypergiants''
with very high infrared luminosities. The effect may become
significant starting from $\mdot \sim 100$.

Further understanding of supercritical accretor winds 
will require methods used for
stellar atmosphere calculations. Rosseland mean for $n \lesssim
10^{10} \cmc$ and $T \sim 10^4 \div 10^5\rm K$ is very close to
$\kappa_T$ \citep{OPAL} but in certain spectral ranges the wind should be
less transparent. 
An edge-on ULX will mimick an OB-hypergiant with dense and
fast wind or a low-temperature hydrogen-rich WR.

\subsection{Photoionized Nebulae}\label{sec:nublado}

\begin{figure}
\center{
\FigureFile(0.7\columnwidth,\texthight){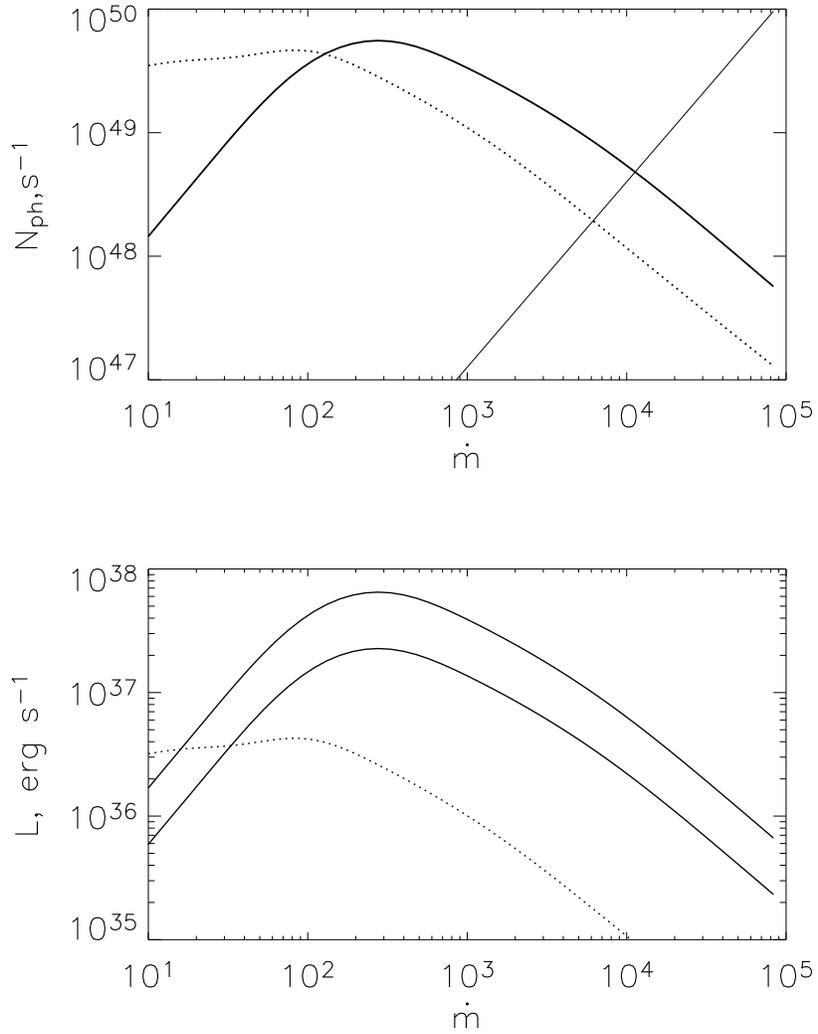}
\caption{
Numbers of H (solid curve) and He$^+$ (dotted) -ionizing photons produced by 
our supercritical wind model (upper panel) and the luminosities of relevant
recombination lines (b): H$\alpha$ and H$\beta$ (solid) and
HeII$\lambda$4686 (dotted).  $\alpha=0$, $r_{in} = 1/ 6 \mdot $ and
$\theta_f = 23^\circ$ is assumed. Thin straight
solid line shows the number of quanta absorbed in the wind. \label{fig:ioph0}
}
}
\end{figure}

In the case of highly supercritical accretion
UV and EUV spectra may give much more information about the mass
ejection rate than the X-ray properties. In figure \ref{fig:ioph0}
we show the dependence of
H and He$^+$ -ionising fluxes ($1\div 1.5$ and $4\div 20\,\Ry$ ranges,
respectively) and corresponding luminosities of
recombination emission lines on mass accretion/ejection rate. Here we 
suggest that the HII regions have fixed temperature $T = 10^4\,\rm K$ and density $n=100\,\cmc$. 
Atomic data were taken from \citet{osterbrock}.
One may see in the figure that the high emission-line luminosities
observed in ULX nebulae may be well explained by high EUV luminosities of
the central sources. 
The HeII$\lambda$ 4686 /  H$\beta$ ratios predicted by the model are 
close to the high HeII$\lambda$4686 /  H$\beta$ ratios $\sim 0.2$
measured for some high-excitation ULX nebulae or the inner
high-excitation parts observed in some of the ULX shells \citep{exart}.

The situation becomes more complicated if one takes into account
absorption in the wind optically thin to electron scattering. 
Density at the wind electron-scattering photosphere is

\begin{equation}
n = \frac{\dot{M}}{\Omega_f R_{out}^2 v } \simeq
3\times 10^9 
\mdot_3^{-3/2} 
 \cmc
\end{equation}
\noindent
Recombination will occur in a layer having thickness:
\begin{equation}
 \begin{array}{l}
  \Delta R = \frac{v}{a n} \simeq \\
  \qquad{} \simeq 2 \times 10^{11} 
  \mdot_3 
M_{10} 
  T_4^{-1/2} \rm cm, \\
 \end{array}
\end{equation}
\noindent
where $a \simeq 2.6\times 10^{-13}\, T_4^{-1/2} \rm cm^3 s^{-1}$ is the effective recombination
rate and $T_4$ is gas temperature in $10^4\,\rm K$ units. The size of the
recombination region is close to or smaller than the size of the outer
photosphere. 
Actually that means that the outer photosphere of the wind will be
optically thick to the EUV radiation of the central source. 

Hard radiation from the pseudo-photosphere may however ionize the
gas. Because the
density of the wind falls off rapidly two regimes appear: ionized and
neutral wind. First case makes the photosphere of the wind a UV object
with high bolometric correction and a compact HII-region coincident
with the X-ray source. In the latter the photosphere is much cooler
and mimicks a hypergiant with broad emission lines.

The ability of the central source to ionize the wind may be calculated
as follows. Recombination rate integrated over the outer wind is
\begin{equation}
I =\int_{R_{out}}^{R_{max}} a \Omega_f n^2(R) R^2 dR
\end{equation}
\noindent
$R_{max}$ here is the radius of Str\"omgren zone ionized by the
central source. 
$I$ may be used as an estimate for the number of hydrogen-ionizing quanta intercepted by
the wind, if $R_{max}$ is set to infinity:
\begin{equation}
S \simeq \frac{2}{\cos^{3/2}\cos\theta_f} \frac{a R_G}{\sigma_T^2}
\mdot^{3/2} \simeq 1.2\times 10^{47} \mdot_3^{3/2} \rm \, s^{-1}
\end{equation}
\noindent
This recombination rate is usually lower than the quanta production
rate, affecting only the highest accretion rates.
In figure \ref{fig:ioph0} we present the ionizing quanta production
rates for different mass ejection rates. 
ULX nebulae may appear even brighter due to two additional effects: ``stripping'' of the
outer wind photosphere (the effect proposed in previous section) and
higher accretion power due to $K < 1$.

\section{Conclusions}

Optically thick wind with irradiation is capable to
explain the SEDs of ULXs in the standard X-ray band and even the
high-energy curvature that is difficult to explain in the framework of
unsaturated comptonisation cool-disc IMBH model \citep{curva}.
High-energy cut-off is predicted to appear at several \keV. Higher
observed values of $T_{in} \sim 1\div 2\,\keV$ may be
explained by applying a hardening factor $\sim 1\div 3$ similar to
those predicted for accretion discs in X-ray binaries. 

Outcoming spectra observed at low inclinations are similar to
the spectra of slim discs and resemble {\it p-free} model spectra with
the $p$ parameter close to $0.5$.
X-ray spectra of known ULXs are approximated equally good by our
model, {\it p-free} and standard disc + power law two component
models. 
However the parameters are poorly constraint supporting the
idea that the properties of the X-ray spectrum depend rather weakly on
the accretion disc and wind parameters. Relatively high inner temperatures argue for
the funnel interior to be transparent. 
In this assumption, $r_{in}$ parameter may be used to estimate the
mass ejection rate that appears to be of the order $\mdot \sim 100\div
300$ for the two sources analysed. 

We stress the extreme importance of irradiation effects providing
mild geometrical collimation of the observed X-ray radiation. In a
simple assumption of local absorption and re-radiation of the absorbed
energy we show that the temperature of the funnel wall surface is
altered by about 20\% in the inner parts of the funnel, and the
outcoming apparent X-ray luminosity becomes about $2\div 3$ times higher. 


Photoionized nebulae are likely to be formed around supercritical accretors. Ionizing
quanta production rates suggest that in most cases a supercritical
accretor is capable to produce a photoionized HII-region with
bright optical emission line luminosities $\sim 10^{37}\ergl$ but higher
luminosities may appear as well.


\bigskip

The work was partially supported by the RFBR/JSPS grant 05-02-19710.


\begin{table}
\caption{Best fitting results for the two selected ULXs. Errors correspond to
  90\% confidence range. }\label{tab:xspectra}
\center{ 
\begin{tabular}{rcc}
\hline
\hline
                                            &       NGC4559 X-7                  &        NGC6946 ULX-1 \\
\hline
\noalign{\medskip}   
             \multicolumn{3}{c}{standard disc + power law}
	     \\
\hline
    $N_H,\, 10^{22}\rm cm^{-2}$                &   $0.219^{0.03}_{-0.02}$     & $0.44^{+0.08}_{-0.05}$  \\
    $T_{in}, \keV$                           &   $0.155^{+0.010}_{-0.009}$    &  $0.156^{+0.018}_{-0.019}$ \\
     standard disc normalization                          &   $150^{+120}_{-60}$         &   $620^{+2000}_{-200}$            \\
    $\Gamma$                                &     $2.24^{+0.05}_{-0.05}$  &      $2.46^{+0.15}_{-0.09}$    \\
    power law normalization                         &   $2.3^{+0.14}_{-0.14} \times 10^{-4}$   &  $3.3^{+0.6}_{-0.5} \times 10^{-4}$ \\
\noalign{\smallskip}
    $\chi ^2 / DOF$                               &     $648/673$                      &      $523/504$      \\
\noalign{\smallskip}
    $L_{model}, 10^{39}\ergl$                     & $9.6^{+0.4}_{-1.6}$        &       $3.1^{+3}_{-1.0}$        \\
\noalign{\bigskip}
\hline
             \multicolumn{3}{c}{p-free disc} \\
\hline
    $N_H,\, 10^{22} \rm cm^{-2}$                &        $0.0407^{+0.0011}_{-0.0009}$   &  $0.293^{+0.03}_{-0.016}$  \\
    $T_{in}, \keV$                           &           $
    0.64^{+0.22}_{-0.23}$     &   $ 2.3^{+0.4}_{-0.4} $   \\
    $p$                                     &      $ 0.460^{+0.004}_{-0.005}$ &  $ 0.404^{+0.01}_{-0.01} $  \\
    p-free disc normalization                           &     $1.4^{+0.9}_{-0.5} \times 10^{-4}$ &  $3.3^{+2}_{-0.4} \times 10^{-4}$ \\
\noalign{\smallskip}
    $\chi ^2 / DOF$                           &                 $734/674$       &   $665/509$  \\
\noalign{\smallskip}
    $L_{model}, 10^{39}\ergl$              &    $9^{+6}_{-3}$     & $3\pm1$    \\
\noalign{\bigskip}
\hline        
             \multicolumn{3}{c}{self-irradiated multi-color funnel}\\
\hline
 $N_H,\, 10^{22}\rm  cm^{-2}$        & $0.167^{+0.026}_{-0.015}$    & $0.293^{-0.016}_{+0.011}$ \\
$T_{in},\keV$                    &     $1.41^{+0.16}_{-0.12}$       & $1.8^{+0.3}_{-0.2}$ \\
$r_{in}$             &   $1.6^{-0.9}_{+1.1}\times 10^{-3}$ & $6^{+5}_{-3}\times 10^{-4}$ \\
$\theta,\deg$   &   $12.2^{+0.7}_{-0.6}$  &   $19^{+2}_{-2}$ \\
    normalization  &      $30^{+62}_{-21}$          &       $74^{+50}_{-50}$               \\
\noalign{\smallskip}
    $\chi ^2 /DOF$   &       $660/673$       &       $604/504$             \\
\noalign{\smallskip}
    $L_{model}, 10^{39}\ergl$      &          $10.5^{+6}_{-4}$    & $3^{+2}_{-1}$   \\
\noalign{\smallskip}
\hline
\noalign{\smallskip}
\end{tabular}}
\end{table}


\begin{appendix}

\section{Integration over $\varphi$}\label{sec:app:phint}

The integral over $\varphi$ can be calculated as follows:

$$
I = R^2 \int_{-\pi}^{\pi} w(R^\prime / R, \theta_f, \varphi)d\varphi = 
\frac{1}{2\pi} x \sin^2 \theta_f \cos^2 \theta_f \int_{-\pi}^{\pi} \frac{(1-\cos\varphi)^2}{\left( a+b \cos\varphi \right)^2}d\varphi
$$

Where $a=1-2x \cos^2 \theta_f +x^2$, $b=-2x \sin^2 \theta_f$. 

$$
\begin{array}{l}
 I= \frac{1}{2\pi} x \sin^2 \theta_f \cos^2 \theta_f \left( \int_{-\pi}^{\pi} \frac{1}{\left( a+b \cos\theta_f \right)^2}d\varphi -
2 \int_{-\pi}^{\pi} \frac{\cos \varphi }{\left( a+b \cos\theta_f \right)^2}d\varphi + 
\int_{-\pi}^{\pi} \frac{\cos^2 \varphi }{\left( a+b \cos\theta_f \right)^2}d\varphi \right)=\\
 \qquad{} = \frac{1}{2\pi} x \sin^2 \theta_f \cos^2 \theta_f \left( \left(1-\frac{a^2}{b^2}\right) \int_{-\pi}^{\pi} \frac{1}{\left( a+b \cos\theta_f \right)^2}d\varphi -
2 \left(1+\frac{a}{b}\right) \int_{-\pi}^{\pi} \frac{\cos \varphi }{\left( a+b \cos\theta_f \right)^2}d\varphi + 
\frac{2\pi}{b^2} \right)\\
\end{array}
$$

The two integrals can be expressed as follows (for details see for example \cite{dwight} or any other table of integrals):

$$
\begin{array}{l}
\int_{-\pi}^{\pi} \frac{1}{\left( a+b \cos\theta_f \right)^2}d\varphi = \frac{a}{\left(a^2-b^2\right)}
\int_{-\pi}^{\pi} \frac{1}{ a+b \cos\theta_f}d\varphi = \frac{2\pi a}{\left(a^2-b^2\right)^{3/2}}\\
\int_{-\pi}^{\pi} \frac{\cos \varphi }{\left( a+b \cos\theta_f \right)^2}d\varphi = -\frac{b}{a^2-b^2}
\int_{-\pi}^{\pi} \frac{1}{ a+b \cos\theta_f}d\varphi = -\frac{2\pi b}{\left(a^2-b^2\right)^{3/2}}\\
\end{array}
$$

Finally, the integral value becomes:

$$
\begin{array}{l}
I = \frac{ x \sin^2\theta_f \cos^2\theta_f}{b^2} \left( 1-
\sqrt{\frac{a+b}{a-b}} \frac{a-2b}{a-b} \right) = \\
\frac{2\pi x \cot^2\theta_f}{4} \left( 1- \frac{|1-x|}{\sqrt{1-2x \cos(2\theta_f) +x^2}} \frac{1-2x (1-3 \sin^2 \theta_f) +x^2}{1-2x \cos(2\theta_f) +x^2} \right)\\
\end{array}
$$

\end{appendix}


\newpage

\begin{thebibliography}{}

\bibitem[Abell \& Margon(1979)]{ss_cinema}
Abell, G. O. \& Margon, B. 1979, Nature, 279, 701

\bibitem[Abolmasov et al.(2007)]{exart}
Abolmasov, P., Fabrika, S., Sholukhova, O. \& Afanasiev, V. 2007,
Astrophysical Bulletin, 62, 36

\bibitem[Abolmasov et al.(2008)]{mf16}
Abolmasov, P., Fabrika, S., Sholukhova, O. \& Kotani, T. 2008,
submitted to PASJ; arXiv:0809.0409


\bibitem[Abramowicz et al.(1978)]{abram78}
Abramowicz, M. A., Jarozhy\'nski, M. \& Sikora, M. 1978, A\&A, 63, 221

\bibitem[Abramowicz et al.(1980)]{agn_tori}
Abramowicz, M. A., Calvani, M. \& Nobili, L. 1980, ApJ, 242, 772

\bibitem[Abramowicz(2004)]{abram2004}
Abramowicz, M. A. 2004, Invited lecture at the conference ``Growing
Black Holes: Accretion in a Cosmological Context'' (Garching, Germany,
21-25 June 2004) In print: ``Growing Black Holes'', Eds. A. Merloni,
S. Nayakshin and R. Sunyaev, ``ESO Astrophysics Symposia Series'',
Springer-Verlag, Berlin, 2004 ; astro-ph/0411185

\bibitem[Bauer \& Brandt(2004)]{ic10_bauer}
Bauer, F. E. \& Brandt, W. N. 2004, ApJL, 601, 67

\bibitem[Blair et al.(2001)]{BFS}
Blair, W. P., Fesen, R. A. \& Schlegel, E. M. 2001 The Astronomical Journal, 
121, 1497

\bibitem[Blundell et al.(2001)]{blundell2001}
Blundell, K. M., Mioduszewski, A. J. \& Muxlow, T. W. B. 2001, ApJL,
562,  79

\bibitem[Borozdin et al.(1999)]{bor99}
Borozdin, K., Revnivtsev, M., Trudolyubov, S., Shr\"ader, C. \&
Titarchuk, L. 1999, ApJ, 517, 367


\bibitem[Clarkson et al.(2003)]{superorb}
Clarkson, W. I., Charles, P. A., Coe, M. J., Laycock, S., Tout,
M. D. \& Wilson, C. A. 2002, MNRAS, 339, 447


\bibitem[Cropper et al.(2004)]{ngc4559_sp}
Cropper, M., Soria, R., Mushotzky, R. F.,  Wu, K. Markwardt, C. B. \&
Pakull, M. 2004, MNRAS, 353, 1024

\bibitem[Dewangan et al.(2005)]{curva}
Dewangan, G. C., Griffiths, R. E. \& Rao, A. R. 2005,
astro-ph/05111102

\bibitem[Dolan et al.(1997)]{dolan}
Dolan J.~F., Boyd P.~T., Fabrika S. \etal
1997 A\&A, 327, 648

\bibitem[Dwight(1961)]{dwight}
Dwight, H. B. 1961, ``Tables of Integrals'', New York: the MacMillan
Company

\bibitem[Eggum et al.(1988)]{EGK}
Eggum, G.E.,  Coroniti, F.V. \& Katz, J.I. 1985, \apj, 330, 142

\bibitem[Fabbiano(1989)]{fabbiano88}
Fabbiano, G., 1989, ARA\&A, 27, 87

\bibitem[Fabrika(2004)]{ss2004}
Fabrika, S.  ``Supercritical disk and jets of SS433''
2004, ASPR, vol. 12


\bibitem[Frank et al.(2002)]{accretion_power}
Frank, J., King, A. \& Raine, D. 2002, ``Accretion Power in
Astrophysics'', Cambridge: Cambridge University Press, third edition


\bibitem[Gierli\'nski \& Done(2004)]{done}
Gierli\'nski, M. \& Done, C. 2004, MNRAS Letters, 349, 7 

\bibitem[Gon\c{c}alves \& Soria(2006)]{goncalves}	
Gon\c{c}alves, A. C. \& Soria, R. 2006, MNRAS Letters, 371, 673

\bibitem[Goranskij et al.(1998)]{gorss}
Goranskii, V. P., Esipov, V. F. \& Cherepashchuk, A. M. 1998, 
Astronomy Reports, 42, 209

\bibitem[Heinzeller et al.(2007)]{heinz}
Heinzeller, D., Mineshige, S. \& Ohsuga, K. 2007, MNRAS, 372, 1208

\bibitem[Iglesias \& Rogers(1996)]{OPAL}
Iglesias, C. A. \& Rogers, F. G. 1996, ApJ 464, 943

\bibitem[Jahnke \& Emde(1960)]{Q_weisstein}
Jahnke, E. and Emde, F. (revised by F. L\"osch), ``Tables of Higher
Functions'' (McGraw-Hill, New York, 1960).

\bibitem[Kaaret et al.(2006)]{m82_kaa}
Kaaret, P., Simet, M. G. \& Lang, C. C. 2006, ApJ, 646, 174

\bibitem[Katz(1986)]{katz86}
Katz, J. 1986, Comments Astrophys., 11, 201

\bibitem[Kobulnicky \& Fryer(2007)]{KoFr2007}
Kobulnicky, H. A. \& Fryer, C. L. 2007, ApJ, 670, 747

\bibitem[Koz{\l}owski et al.(1978)]{kozl78} 
Koz{\l}owski, M., Jaroshy\'nski, M. \& Abramowicz, M. A. 1978, A\&A,
  63, 209




\bibitem[Lucy(2007)]{lucy}
Lucy, L. B. 2007, A\& A, 457, 629

\bibitem[Martin et al.(2005)]{martingalex}
Martin, C. et al., 2005, ApJ, 619, 1L

\bibitem[Mineshige (1993)]{min93}
Mineshige, S. 1993, Astrophysics and Space Science, 210, 83

\bibitem[Mineshige et al.(1994)]{discpbb}
Mineshige, S., Hirano, A., Kitamoto, S., Yamada, T. \& Fukue, J. 1994,
ApJ, 426, 308

\bibitem[Mitsuda et al.(1984)]{discbb}
Mitsuda, K., Inoue, H., Koyama, K. et al. 
1984, PASJ, 36, 741

\bibitem[Nishiyama et al.(2007)]{nishi07}
Nishiyama, S., Waratai, K.-y. \& Fukue, J. 2007, PASJ, 59, 1227

\bibitem[Ohsuga et al.(2005)]{ohsuga2005}
Ohsuga, K., Mori, M., Nakamoto, T. \& Mineshige, S. 2005, ApJ, 628,
368

\bibitem[Okuda(2002)]{okuda}
Okuda T. 2002, PASJ, 54, 253

\bibitem[Osterbrock \& Ferland(2006)]{osterbrock}
Osterbrock, D. E. \& Ferland, G. ``Astrophysics of Gaseous Nebulae and
Active Galactic Nuclei'' 2006, 2nd. ed. by D.E. Osterbrock and
G.J. Ferland. Sausalito, CA: University Science Books, 2006


\bibitem[La Parola et al.(2001)]{hoix_parola}
La Parola, V., Peres, G., Fabbiano, G., Kim, D. W. \& Bocchino, F. 2001, ApJL, 556, 47


\bibitem[Poutanen et al.(2007)]{poutanen} 
Poutanen, J., Lipunova, G., Fabrika, S., Butkevich, A. \& Abolmasov,
P. 2007, MNRAS, 377, 1187

\bibitem[Roberts(2007)]{roberts_review}
Roberts, T. P. 2007, Astrophysics and Space Science, 311, 203 

\bibitem[Roberts \& Colbert(2003)]{ngc6946_RoCo}
Roberts, T. P. \& Colbert, E. J. M. 2003, MNRAS, 341, 49


\bibitem[Shakura \& Sunyaev(1973)]{ss73}
Shakura, N. I. \& Sunyaev, R. A. 1973, A\&A, 24, 337


\bibitem[Shklovskii(1981)]{sklov81}
Shklovskii, I.S. 1981, Sov. Astron., 25, 315

\bibitem[Soria et al.(2005)]{ngc4559x7_soria}
Soria, R., Cropper, M., Pakull, M., Mushotzky, R. \& Wu, K. 2005,
MNRAS, 356, 12

\bibitem[Stobbart et al.(2006)]{soft_excess}
Stobbart, A.-M., Roberts, T. P. \& Wilms, J. 2006, MNRAS, 368, 397

\bibitem[Turner et al.(2001)]{epic}
Turner, M. J. L., Abbey, A., Arnaud, M., et al. 2001, A\&AL, 365, 27

\bibitem[Van den Heuvel(1981)]{vdh81}
Van den Heuvel, E. P. J. 1981, Vistas in Astronomy, 25, 95

\bibitem[Vierdayanti et al.(2006)]{vier2006}
Vierdayanti, K., Mineshige, S., Ebisawa, K. \& Kawaguchi, T. 2006,
PASJ, 58, 915

\bibitem[Watarai et al.(2005)]{wata}
Watarai, K.-y., Ohsuga, K., Takahashi, R. \& Fukue, J.
2005, PASJ, 57, 513

\end{thebibliography}
\end{document}